\documentclass[]{pasj02} 
\usepackage{url}
\usepackage{tikz}

\jyear{2024}
\Received{}%{yyyy/mm/dd}
\Accepted{}%{yyyy/mm/dd}
\newcommand{\onohiula}[1]{`\={O}nohi`ula }

\begin{document} 

\title{Prime Focus Spectrograph on the Subaru Telescope: Overview of Science Operations}

%%% begin:list of authors
% Do NOT capitalize all letters in "textsc".
\author{
  Masayuki \textsc{Tanaka}\altaffilmark{1,2}\altemailmark\orcid{0000-0002-5011-5178} \email{masayuki.tanaka@nao.ac.jp},
  Akira \textsc{Arai}\altaffilmark{3},
  Wanqiu \textsc{He}\altaffilmark{1},
  Miho N. \textsc{Ishigaki}\altaffilmark{1,2},
  Eric \textsc{Jeschke}\altaffilmark{3},
  Russell \textsc{Kackley}\altaffilmark{3},
  Shintaro \textsc{Koshida}\altaffilmark{3},
  Yuki \textsc{Moritani}\altaffilmark{3,2},
  Masato \textsc{Onodera}\altaffilmark{3,2},
  Vera Maria \textsc{Passegger}\altaffilmark{3},
  Tae-Soo \textsc{Pyo}\altaffilmark{3},
  Yuhei \textsc{Takagi}\altaffilmark{3},
  Naoyuki \textsc{Tamura}\altaffilmark{3},
  Ichi \textsc{Tanaka}\altaffilmark{3},
  Kiyoto \textsc{Yabe}\altaffilmark{3,2},
  Sadman S. \textsc{Ali}\altaffilmark{1},
  Javier Gracia \textsc{Carpio}\altaffilmark{4},
  Maximillian \textsc{Fabricius}\altaffilmark{4},
  Wilfred \textsc{Gee}\altaffilmark{3},
  James E. \textsc{Gunn}\altaffilmark{5},
  Michitaro \textsc{Koike}\altaffilmark{1},
  Arnaud \textsc{Le Fur}\altaffilmark{5},
  Zhuoming \textsc{Li}\altaffilmark{1},
  Yongming \textsc{Liang}\altaffilmark{6},
  Craig \textsc{Loomis}\altaffilmark{5},
  Robert \textsc{Lupton}\altaffilmark{5},
  Sogo \textsc{Mineo}\altaffilmark{1},
  Takahiro \textsc{Morishima}\altaffilmark{1},
  Sakurako \textsc{Okamoto}\altaffilmark{3,2},
  Yuki \textsc{Okura}\altaffilmark{1},
  Paul \textsc{Price}\altaffilmark{5},
  Martin \textsc{Reinecke}\altaffilmark{4},
  Michael A. \textsc{Strauss}\altaffilmark{5}
% B-Firstname \textsc{B-Familyname},\altaffilmark{2}$^{,\dag}$\orcid{0000-0000-0000-0000}
% C-Firstname \textsc{C-Familyname},\altaffilmark{3}\altemailmark \email{ccccc@xxx.xxx.xx.xx}
% and 
% D-Firstname \textsc{D-Familyname}\altaffilmark{2}\altemailmark\orcid{0000-0000-0000-0000} \email{ddddd@xxx.xxx.xx.xx}
}
\altaffiltext{1}{National Astronomical Observatory of Japan, 2-21-1 Osawa, Mitaka, Tokyo 181-8588, Japan}
\altaffiltext{2}{Graduate Institute for Advanced Studies, SOKENDAI, 2-21-1 Osawa, Mitaka, Tokyo 181-8588, Japan}
\altaffiltext{3}{Subaru Telescope, National Astronomical Observatory of Japan, 650 North A'ohoku Place, Hilo, HI 96720, USA}
\altaffiltext{4}{Max-Planck-Institut fuer extraterrestrische Physik, Giessenbachstrasse, D-8574 Garching bei Muenchen, Germany}
\altaffiltext{5}{Department of Astrophysical Sciences, Princeton University, 4 Ivy Lane, Princeton, NJ 08544, USA}
\altaffiltext{6}{Institute for Cosmic Ray Research, The University of Tokyo, 5-1-5 Kashiwanoha, Kashiwa, Chiba 277-8582, Japan}

%\footnotetext[$\dag$]{Present address: ....}

%%% end:list of authors

%% !!! Select 3 to 5 words from PASJ's key words !!! 
%% List of Key Words: https://academic.oup.com/pasj/pages/Pasj_Keywords 
%% "\KeyWords{ }" always has to be placed before ``\maketitle'' 

\maketitle

\begin{abstract}
  The paper presents the science operation framework for Prime Focus Spectrograph (PFS or \onohiula\ in its Hawaiian name)
  installed
  at the 8.2m Subaru Telescope on the summit of Maunakea. PFS is a massively multiplexed, wide-field, fiber-fed spectrograph
  covering 1.25 square degrees with 2386 science fibers. The instrument has been offered to the Subaru scientific community
  since March 2025. In order to fully exploit the unique capabilities of PFS, the Subaru Telescope
  has introduced a new, dedicated science operation framework for PFS. The default observing mode is
  queue observing, and multiple observing programs (in the same field) can be executed in the same exposure to
  achieve high observing efficiency.
  The quality of an exposure is based on the delivered signal-to-noise ratio and is quantified in terms of 'effective
  exposure time', and exposures are taken until the allocated 'fiber hours' for each program or target are achieved.  
  The fiber hour is a new unit for observing time at Subaru; if we expose a fiber for 1 hour under the fiducial conditions,
  it is 1 fiber hour. Each observing program is granted the total fiber hours by the Time Allocation Committee.
  In addition to normal observing programs,
  which are selected through the standard peer-review process, there are two filler categories; community
  filler and observatory filler. As the names imply, the former is proposed by the community and the latter is prepared by the observatory.
  These filler targets are used whenever unassigned fibers are available. After an observing run, the data are
  fully reduced by the observatory and delivered to the users through the PFS Science Platform, a cloud-based data
  analysis environment. The paper gives a summary of all of this new framework and the actual
  implementation of it. We emphasize that the new framework is built upon close iterations that have taken place over several years
  between the observatory and the community.
  We continue to make improvements, and the reader is referred to the Subaru
  Telescope website for the latest information.
\end{abstract}

%\pagewiselinenumbers 

%--------------------------------------------------------------------------------------------------------
%--------------------------------------------------------------------------------------------------------
\section{Introduction}
\label{sec:introduction}

%\noindent IMPORTANT NOTICE\\
%1. Manuscript for submission must be in the same format as a published papers. \\
%2. Line numbers should be added to the manuscript. \\
%3. Do NOT use ``\verb|\def|, \verb|\renewcommand|''.\\
%4. Do NOT redefine commands provided by pasj02.cls.  
% Use \verb/\timeform{ }/ for celestial coordinates as shown in the example below.
% (RA, Dec)$_{\rm J2000.0}$ = (\timeform{1h23m45.67s}, \timeform{6D54'32.1''})

Electromagnetic waves are one of the primary probes of celestial objects.
There are different physical processes at work on different physical scales in the Universe
and each process has its own characteristic spectral features. Detailed spectroscopic observations
of objects thus provide us with valuable clues to infer their physical nature.
Emission and absorption lines due to specific atoms/molecules are of particular importance because
they allow us to measure redshifts or radial velocities of objects as we know their precise rest-frame
wavelengths/frequencies.
Radial velocities of stars are crucial to infer their orbital motions and constrain the Galactic dynamics.
Precise radial velocities may even reveal exoplanets orbiting around a star.
Detailed analyses of stellar absorption lines allow us to infer, e.g., the elemental abundances of stars.
For objects at cosmological distances, redshifts yield
distances by assuming a cosmological model, making it possible for us to translate observables such as angular
sizes and apparent fluxes into physical sizes and luminosities to understand objects’ physical nature.
Redshifts are also crucial to constrain cosmological models as well.
These are just examples, and spectroscopy is crucial in many other areas. Spectroscopy has played
a key role in improving our understanding of the Universe for more than a century.

A number of large spectroscopic surveys have been carried out in recent decades. Among them,
the Sloan Digital Sky Survey (SDSS; \cite{york00}) was a pioneer in terms of both quantity
and quality and it revolutionized our understanding of the Universe.
Using a dedicated 2.5m telescope \citep{gunn06},
SDSS imaged about one-third of the celestial sphere in five photometric bands using a wide-field imaging
camera \citep{gunn98} and has collected spectra of millions of stars, galaxies and quasars
in its ongoing survey.
The data were reduced with sophisticated pipeline software, yielding high-quality data for
scientific analyses.
Thanks to the homogeneous high quality imaging and spectroscopy of objects over a wide area to
redshifts of a few tenths, SDSS has firmly anchored our physical understanding of the local universe.
However, statistical properties of objects in the more distant Universe have been left largely
unexplored due to lack of sensitivity. Successor surveys are improving the situation
as we discuss below.
SDSS also demonstrated the importance of combined imaging and spectroscopic observations;
these observing modes are complementary to each other and it is crucial to do both.
Many of the modern observing campaigns indeed utilize both modes
to address some of the most outstanding questions, e.g., weak-lensing cosmology using
high-quality imaging data supplemented by a reference sample of spectroscopic redshifts
of background objects, and later with a more extensive spectroscopic
survey of the cosmic large-scale structure.

We are living in the golden era of wide-field surveys.
The Dark Energy Spectroscopic Instrument (DESI; \cite{desi16a,desi16b}) has been spectroscopically
surveying a wide area of the sky to address the nature of dark matter and dark energy.
The Gaia mission \citep{gaia16} scanned the whole sky over several years and produced a huge amount of
high quality positional information of Galactic stars.
The Euclid satellite is in orbit
and is carrying out its wide-field imaging and spectroscopic mission over a significant
fraction of the sky \citep{euclid25}. The Vera C. Rubin Observatory \citep{ivezic19} is starting
its imaging survey and is going to deliver an unprecedented view of the transient Universe in
the optical wavelengths.
The Roman Space Telescope \citep{akeson19} will be launched soon to carry out
a wide-area imaging and spectroscopic survey in the near-infrared wavelengths.
Massive ground-based spectroscopic surveys such as 4MOST \citep{dejong19} and MOONS \citep{cirasuolo20}
are also coming up. These ongoing and upcoming surveys will undoubtedly improve our understanding of the Universe.

In part as a precursor to these projects, the Japanese astronomical community
started a wide-field imaging survey with Hyper Suprime-Cam (HSC Subaru Strategic Program
or HSC-SSP for short; \cite{aihara18a,aihara18b,miyazaki18,aihara19,aihara22}) mounted on the Subaru Telescope in 2014.
HSC is the most efficient imaging camera (in terms of \'{e}tendue = light collecting area $\times$ field of view)
in the northern hemisphere as of this writing and has made significant
contributions to various astrophysical questions from solar system bodies to cosmology
(see for instance the PASJ special issue in 2018; \cite{aihara18a}). The Subaru telescope is now equipped with a spectroscopic
capability with a similarly wide field of view; the Prime Focus Spectrograph (PFS;
\cite{tamura22,tamura24})
uses the same prime focus optics as HSC, covering a 1.25 deg$^2$ field of view with $\sim2,400$ fibers,
making it possible to do a combined photometric and spectroscopic survey of the sky over
a wide area at the depth of an 8m-class telescope for the first time.
A special aspect of PFS, among other Subaru instruments, is that it is given a Hawaiian name,
`\={O}nohi`ula, which embodies the idea of perceiving the realm of our origins.
This illustrates Subaru’s close relationship and respect to the Hawaiian community.

While the instrument is offered to the general astronomy community, the PFS collaboration \citep{tamura24}
has been awarded 360 nights of observing time over the next $\sim6$ years as a Subaru Strategic Program (SSP).
PFS-SSP is the largest observing program ever approved at Subaru.
The survey is to carry out extensive cosmology, galaxy evolution and galactic archaeology surveys built
upon the imaging data from HSC to address the key questions in dark matter and dark energy.
The science overview is summarized in \citet{takada14}, and more detailed science cases for
the galaxy evolution survey can be found in \citet{greene22}.
The galactic archaeology science is detailed in \citet{chiba26}.
The PFS-SSP survey started in 2025, and intensive data analyses are underway.

Compared to previous Subaru instruments,
PFS is significantly more complex both mechanically and operationally.
The previous science operation framework at Subaru is unable to run PFS in an efficient way as we discuss in detail below.
In order to operate the instrument efficiently and maximize the science output, we have built a new
science operation framework dedicated for PFS. This paper gives an overview of this new framework. We emphasize that the paper presents
only the current snapshot as of April 2026, and the framework will undoubtedly evolve in the future to further optimize the overall
operation. The reader is referred to the PFS instrument page at the Subaru website\footnote{
\url{https://subarutelescope.org/Instruments/PFS/index.html}
}
for the latest information.

The paper is one of the first technical papers from the PFS collaboration. A number of upcoming papers in preparation
include the instrument summary paper, more focused papers on some of the key sub-systems,
data reduction software, and fiber assignment software.
The current paper includes only the high-level summary of the hardware and software and
defers detailed descriptions to these upcoming papers. We note that the goal of
the paper is to summarize the science operations, which deal with not just the SSP survey
but other open-use programs as well.
We therefore attempt to give useful information to general observers.

The paper is structured as follows.
We briefly describe the PFS instrument itself in Section \ref{sec:instrument} and then move on to discuss
the new science operation framework in Section \ref{sec:new_science_operation_framework}.
Section \ref{sec:implementation_overview} summarizes the actual implementation, and
Section \ref{sec:data_distribution} discusses data distribution to the user. Finally,
Section \ref{sec:summary_and_future_improvements} concludes the paper with future prospects.
We adopt the AB magnitude system \citep{oke83} throughout the paper.

%--------------------------------------------------------------------------------------------------------
\section{Instrument Overview}
\label{sec:instrument}

PFS is a fiber-fed, multi-object optical/near-infrared spectrograph installed at the prime focus of the Subaru Telescope \citep{iye04}.
Each fiber is equipped with a positioning device called a 'cobra' with two-axis rotary motors with a small offset between
the two axes, covering a patrol area with a diameter of about 1 arcmin.  Each fiber subtends $\sim1.1$ arcsec on the sky,
although the exact size varies across the focal plane due to optical distortion. Approximately
2,400 fibers are placed in a  hexagonal pattern and are installed at the prime focus to yield
a $\sim1.25$~deg$^2$ diameter field of view \citep{wang22}.
The $60$-meter fibers transmit the light from the prime focus to the spectrographs accommodated in a dedicated,
temperature-controlled clean room inside of the dome enclosure \citep{deoliveira22}.

The spectrograph system covers a wide wavelength range of $380-1260$~nm with four identical spectrograph modules,
each of which consists of three separate cameras (or arms) to cover the blue ($380-650$~nm), red ($630-970$~nm),
and near-infrared ($940-1260$~nm) wavelengths \citep{smee22}.
The blue, red, and near-infrared arms have a resolving power of approximately 2,500, 3,000 and 4,500, respectively.
In addition, the red arm has a medium resolution mode with $R\sim5,500$, covering
a narrower wavelength range of $710-885$~nm.
For the blue and red cameras, pairs of Hamamatsu 2k $\times$ 4k CCDs take $\sim600$ spectra in each spectrograph module,
while Teledyne 4k $\times$ 4k H4RG-15 detectors are used for the near-infrared cameras.
The spectra are placed close to each other on the detectors and the spatial profile of the adjacent spectra
overlap in their wings, which challenges data reduction (see the forthcoming data reduction pipeline paper).
Further details of the instruments can be found in \citet{tamura24} and references therein.

In a PFS exposure, the fibers first move to commanded positions on the focal plane after the telescope
boresight moved to a target position.
During this fiber move, the fibers are back-illuminated (i.e., illuminated from the spectrograph modules)
and the metrology camera installed at
the Cassegrain focus \citep{wang20} looks up the focal plane at the prime focus and measures the positions of the fibers.
Offsets between the target positions and the current positions of the fibers are measured from
metrology camera exposures, and offsets are applied to move the fibers closer to the target positions.
The fibers are repositioned iteratively.  After eight iterations, typically taking three minutes or less,
75\% of the fibers converge within 10$\mu$m (about 0.1 arcsec).
Finally, we complete the field acquisition by setting guide stars using six guide cameras installed just outside of
the fiber arrays and start a sequence of science exposures.

\begin{figure*}
 \begin{center}
  \includegraphics[width=8.8cm]{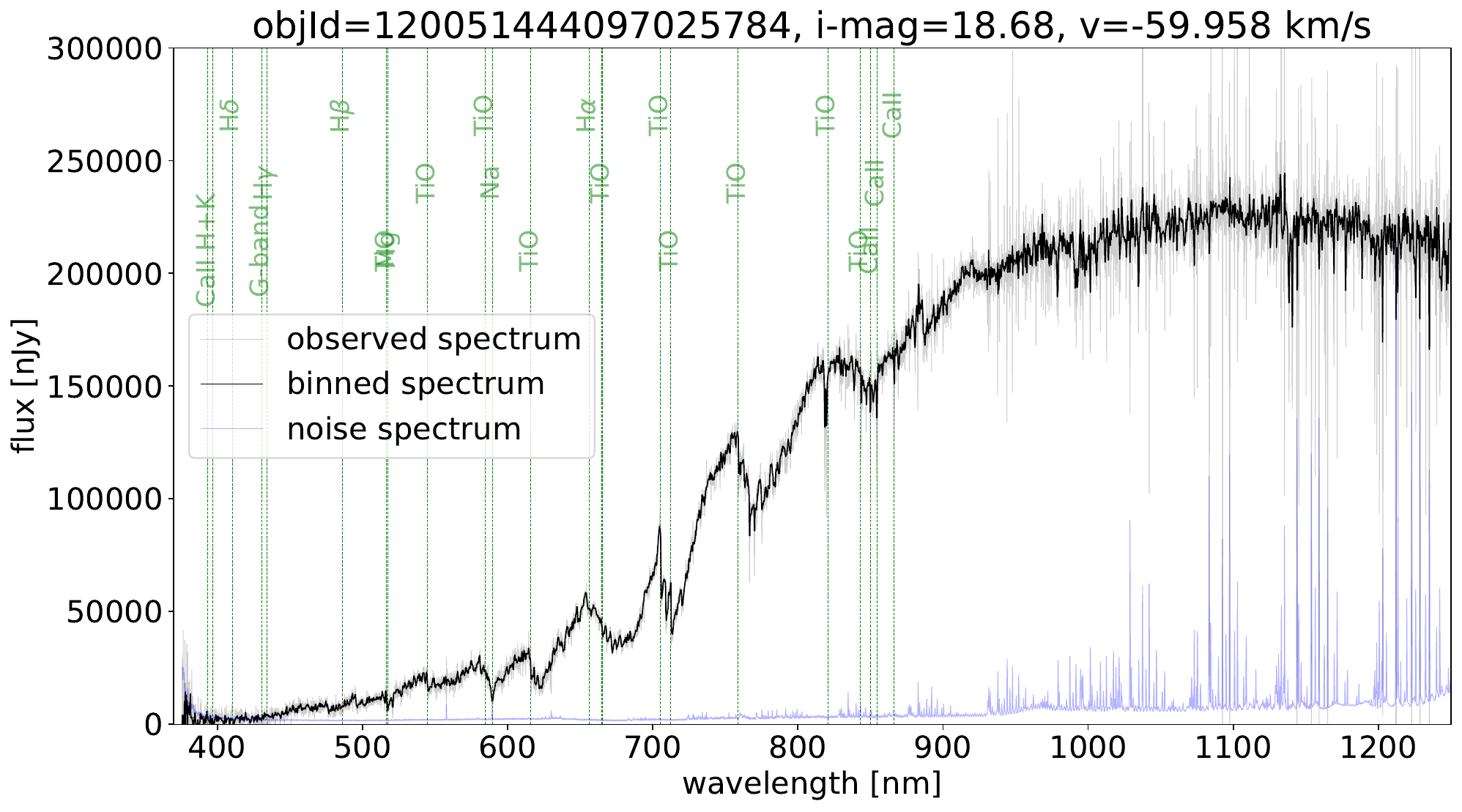}
  \includegraphics[width=8.8cm]{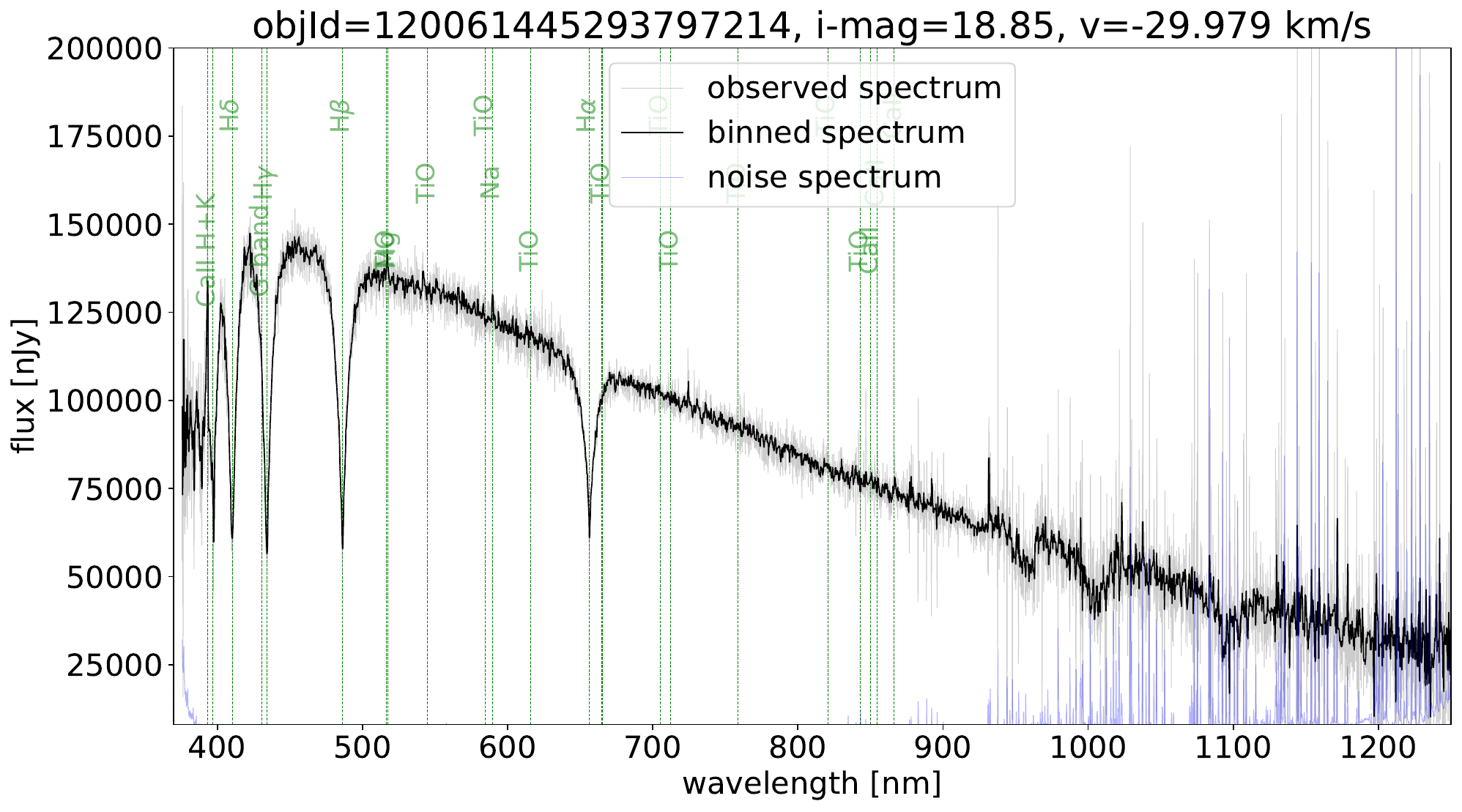}\\\vspace{0.3cm}
  \includegraphics[width=8.8cm]{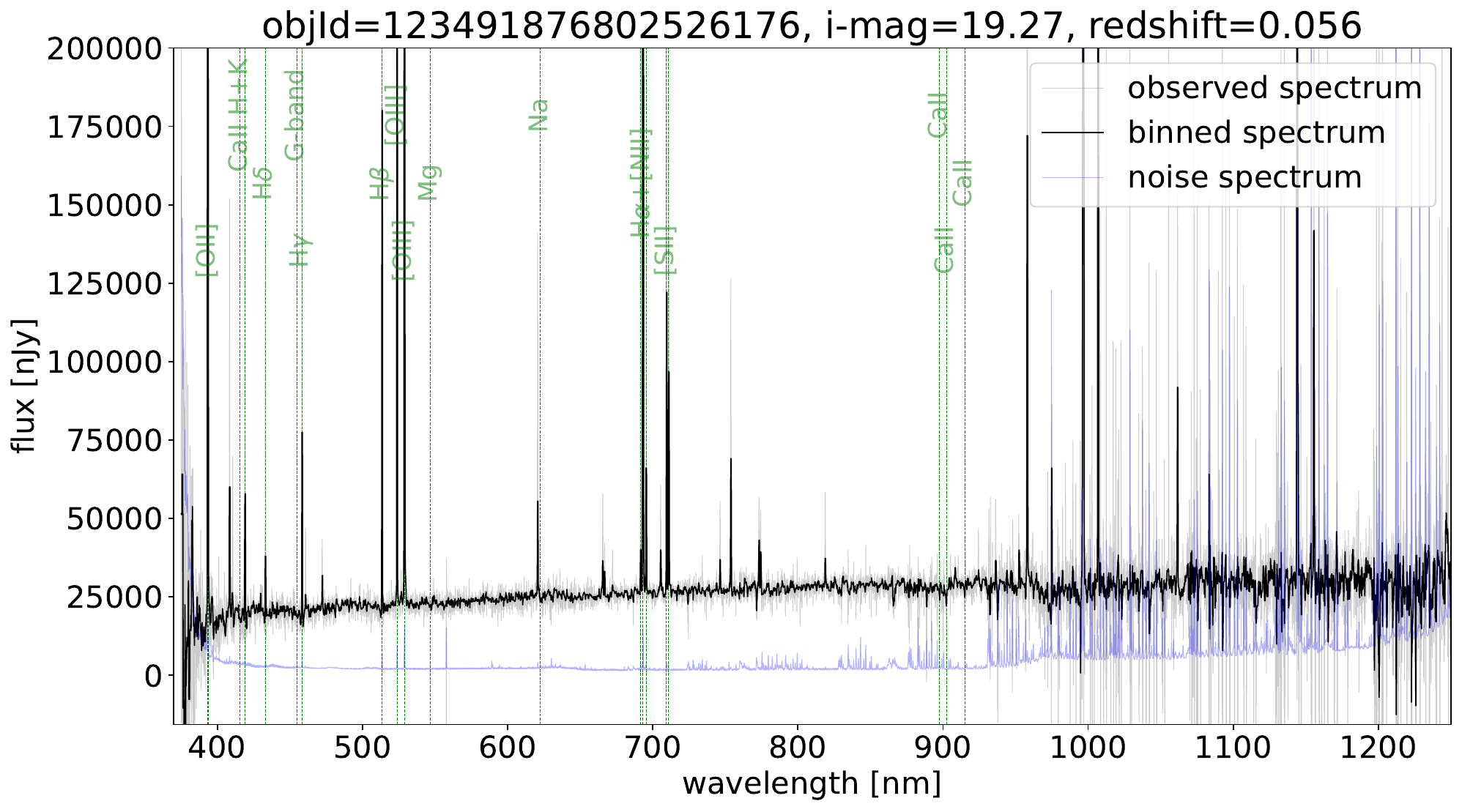}
  \includegraphics[width=8.8cm]{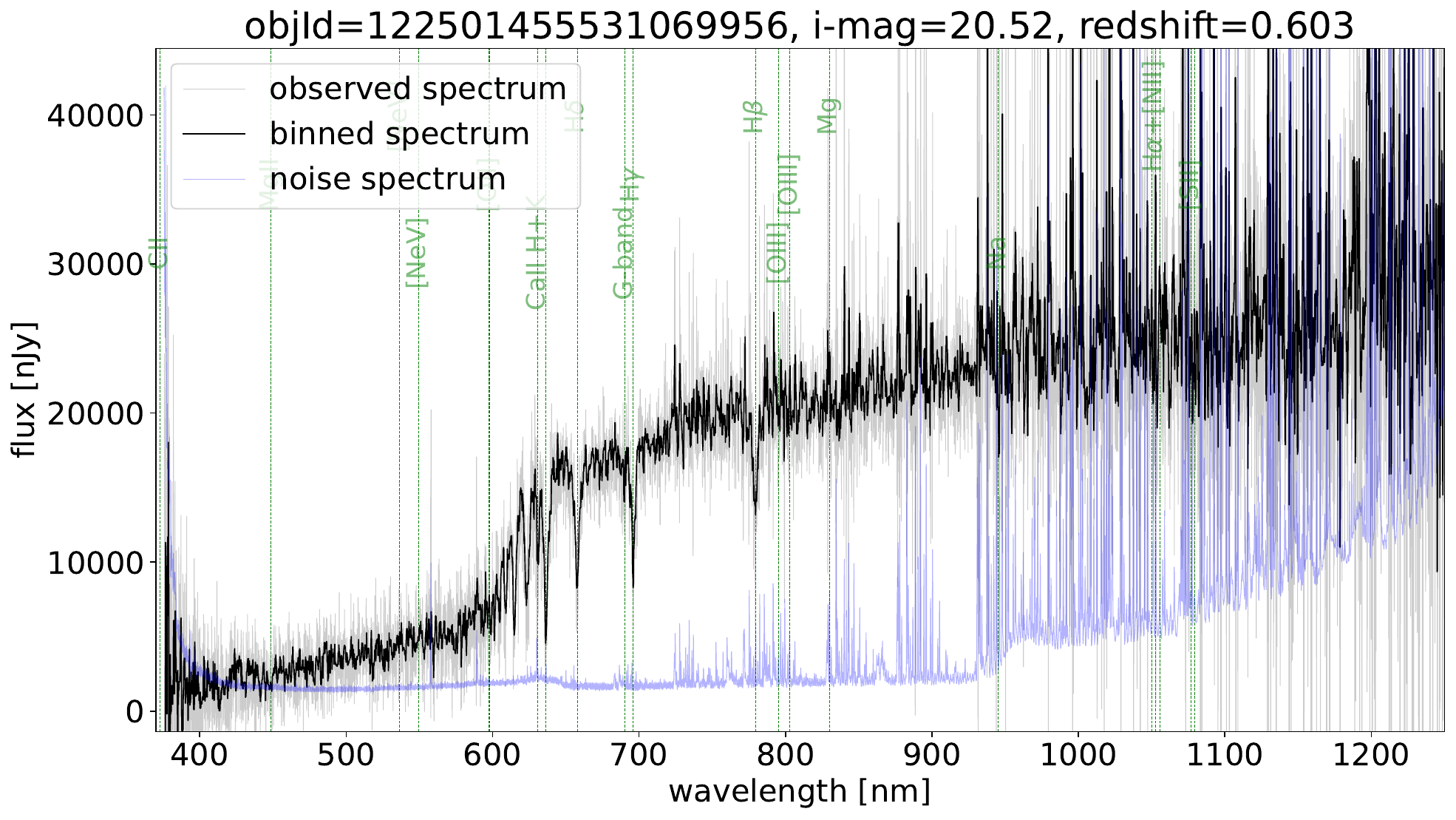}\\\vspace{0.3cm}
  \includegraphics[width=8.8cm]{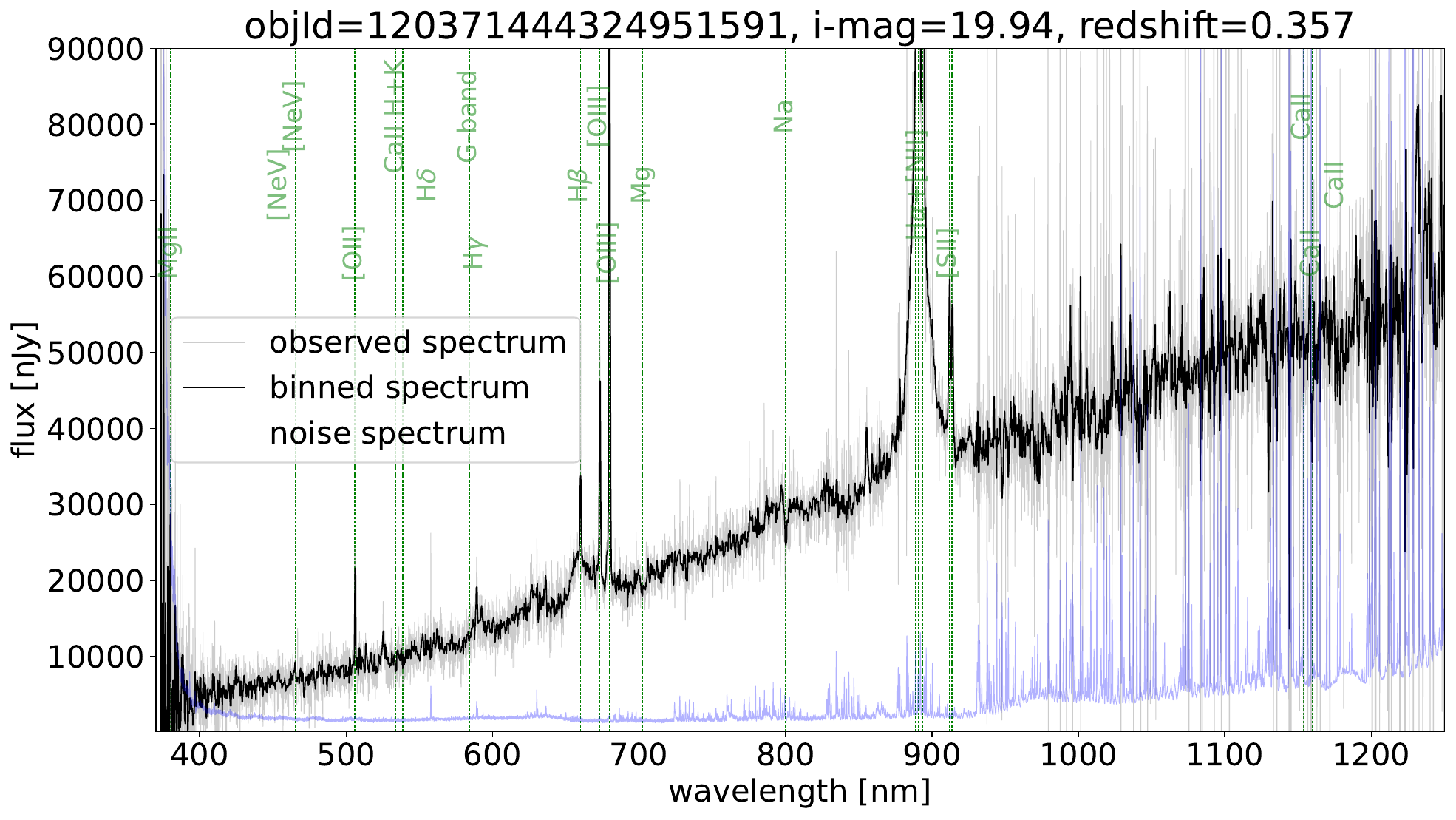}
  \includegraphics[width=8.8cm]{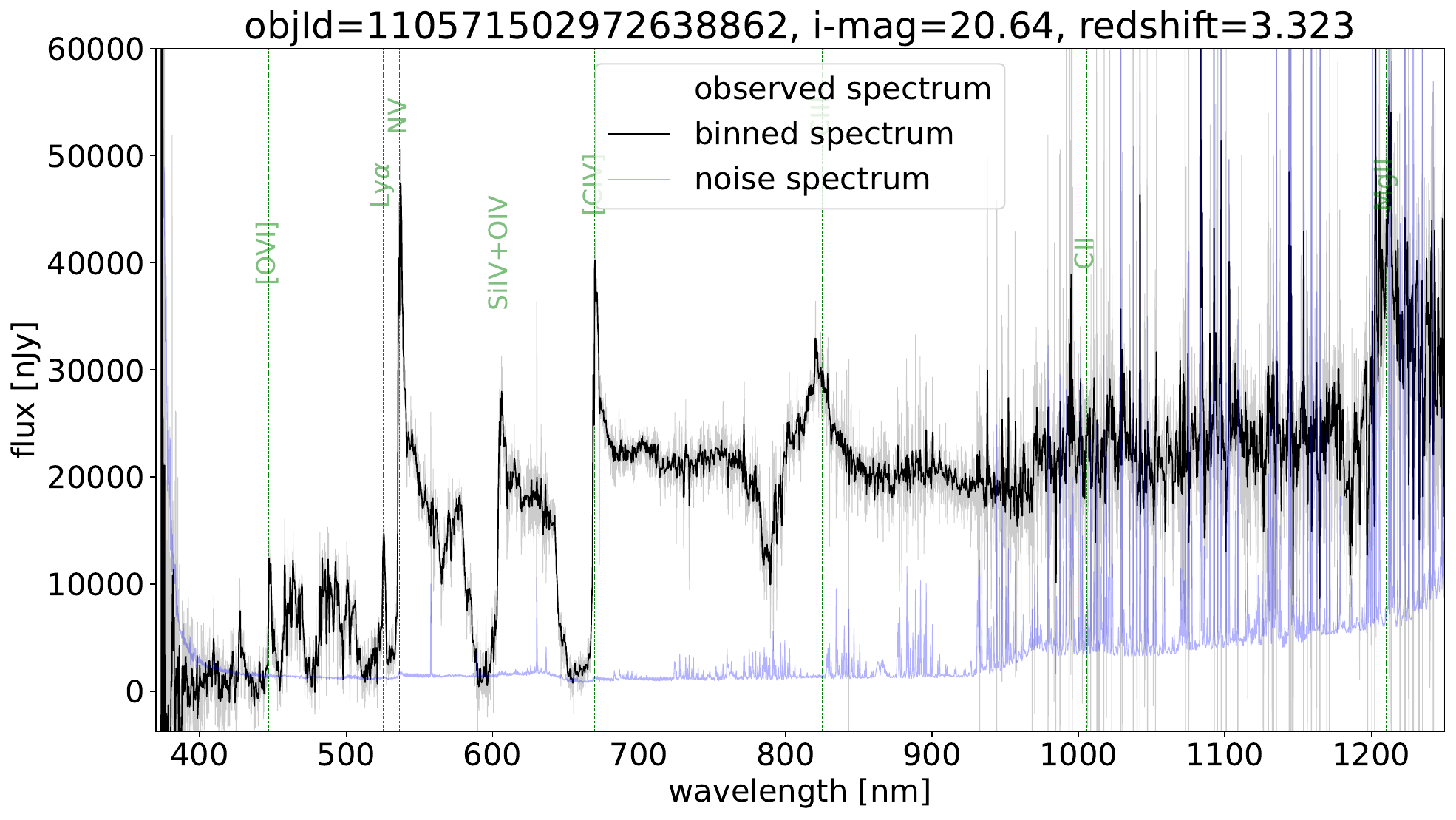}
 \end{center}
 \caption{
   Sample spectra from the first observing run (March 2025).
   The objects shown are observatory filler targets.
   The total exposure time varies from object to object but is typically 30min.
   The gray and black spectra are the original and 10-pixel binned object spectra, respectively.
   The blue spectrum represents the estimated noise per pixel.
   Some of the most prominent lines are indicated by the vertical lines.
   {\bf Top panels:}
   Stellar spectra. About 40\% of observatory filler is stars. Most of them are late-type stars like the left panel,
   but there are early-type stars as shown in the right panel.
   {\bf Middle panels:}
   Roughly 60\% of observatory filler is galaxies. The left is an example of a strong emission line galaxy at
   low redshift and the right is a post-starburst galaxy with pronounced Balmer break.
   {\bf Bottom panels:}
   Quasars comprise only $\sim1$\% of all observatory filler objects, but they are
   interesting objects. The left is a red, dusty quasar. The right is
   a broad absorption line (BAL) quasar.
   {Alt text: A plot with six panels each showing spectral flux in nano Jansky in the vertical axis and
     wavelength in nano meter in the horizontal axis.}
 }
 \label{fig:gallery}
\end{figure*}

Fig. \ref{fig:gallery} is a gallery of spectra of objects observed in the first science operation run\footnote{
In this paper, we define an observing run as a sequence of observing nights with one instrument.
During a PFS run, only PFS is available on Subaru. A PFS run is typically $\sim2$ weeks long centered on the dark time.
} in March 2025.
They are observatory filler targets (see the definition of the observatory filler later in Section \ref{subsec:fillers})
and are available for the community without a proprietary period through the PFS Science
Platform (Section \ref{subsec:pfs_science_platform}). The spectra are typically 30min integration taken
under reasonable conditions and are fully calibrated. The gallery includes
stars, galaxies, and quasars and they demonstrate the quality of the PFS spectra.
Note that the reduction of the near-infrared arm is still premature, and we plan to properly correct for
persistence, which is a memory effect specific to the near-infrared detectors  (e.g., \cite{tulloch19}),
in future processing runs.

%------------------------------

%--------------------------------------------------------------------------------------------------------
\section{New Science Operation Framework}
\label{sec:new_science_operation_framework}

Compared to the current and past spectrographs at Subaru, PFS is by far the most powerful
spectrograph in terms of both multiplexity and field of view. We want to operate the instrument
as efficiently as possible to maximize the science output. The science operation modes at Subaru
are, however, rather limited; most instruments are operated in classical mode, and
an observer has to attend the observation either remotely or onsite. Only HSC can be 
operated in queue mode, although the HSC queue system was designed
primarily for HSC itself and is not readily applicable to other instruments.
There are also so-called ‘service’ observations, which are executed by the observatory staff,
but they are limited up to 4 hours of observing time for one science program and are not suitable for PFS.
We therefore choose to build a new observing framework for PFS from scratch to make the best use of PFS;
this Section summarizes the key concepts. Note that, in what follows, we refer to the science operation
system before PFS as the current system and PFS’s system as the new system.

\subsection{Calibration Objects}

Before we discuss the science operations, let us briefly describe calibration objects because
they are closely related to how we operate the instrument.
In each PFS exposure (or {\tt visit} in the PFS pipeline terminology), we have to dedicate a fraction of
the fibers for calibration objects for pipeline processing. In other words, not all the fibers can be 
used for science targets. The calibration objects include sky objects and flux standards, and
they have to be distributed uniformly across the focal plane for good calibrations.
We briefly describe each type of calibration objects here.
Note that lower level calibrations such as bias, dark, quartz (flat), and arc exposures are taken
by the observatory and are not the responsibility of observers.

\subsubsection{Sky Objects}
\label{subsubsec:sky_objects}

The sky objects sample the blank sky (i.e., sky regions without objects) to obtain the night-sky
spectrum. The sky spectrum is averaged and subtracted from science fibers.
We construct a list of sky objects using the PS1 $3\pi$ survey DR2 \citep{chambers16,flewelling20,magnier20}
and Gaia DR3 \citep{gaia23}.
There is another version of the sky objects based on the HSC-SSP data.
While HSC goes deeper than PS1 by $\sim4$ magnitudes, it covers a significantly smaller area of the sky ($\sim4\%$).
We use the PS1 version by default and the HSC version when needed.

We first use the Gaia DR3 data to mask regions around bright ($<18$th magnitude) stars
using the same method as adopted in HSC \citep{coupon18,aihara22}: we first estimate HSC magnitudes of each
star from the Gaia photometry ($G$, $B_p$ and $R_p$) and define masks. The size of the mask
is different in different filters, but we adopt the largest size to be conservative.
We then mask fainter, extended sources using the PS1 DR1 data. To be specific, we mask
the region within 3 times the Kron radius of each source. To avoid issues with the PS1
measurements such as grossly over-estimated Kron radius due possibly to object blending,
we set the minimum and maximum radius to be 5 arcsec and 100 arcsec, respectively.
Sky regions that are not masked by Gaia nor PS1 masks are considered to be blank sky.
We then draw random positions from the blank sky and generate sky objects at the density of 10 objects per
square arcmin.

We validate the PS1-based sky object catalog using the deeper HSC images. By applying
the Gaia+PS1 mask to a portion of the image from the UltraDeep survey \citep{tanaka17},
which goes down to $\sim27$th magnitude,
the fraction of pixels with a surface brightness of $<25\rm\ mag\ arcsec^{-2}$ is $\sim1\%$,
suggesting that the mask indeed excludes bright objects. Due to the depth of PS1, it is
difficult to exclude very faint sources, but we do not observe a significant continuum level
in the sky objects in our real operations.

\subsubsection{Flux Standards}
\label{subsubsec:flux_standards}

Flux standards also have to be included in each visit for accurate flux calibrations and telluric corrections.
We adopt a similar flux calibration strategy to that adopted in the SDSS;
we observe bright ($17<g<19$) F-stars selected from imaging data, fit models from a stellar spectral
library for accurate stellar typing, use the spectrum with the inferred stellar parameters
as the intrinsic spectrum, and compare the observed and intrinsic spectra to derive
the flux calibration vector. Details of this process can be found in a forthcoming
pipeline paper.

For the standard stars, we primarily use the PS1 catalog; PS1 is sufficiently deep to
identify bright F-stars with $17-19$th magnitudes that are appropriate for flux calibration.
Details of the standard star selection can be found in a forthcoming paper, but in short,
we base our selection on the Gaia DR3 + PS1 DR2 matched catalog with a set of quality cuts applied to
ensure good photometry. We run {\sc brutus} \citep{speagle25}, a stellar parameter inference code,
to evaluate the stellar parameters, in particular the probability distribution of $T_{\rm eff}$. We define
the $T_{\rm eff}$ probability integrated between 6,000 and 7,800K as the F-star probability, and we select
stars with high probabilities in a given field as flux standards.

PS1 covers most but not all of the observable sky from Maunakea, and
we supplement the PS1 flux standards with those selected from Gaia DR3.
We use the temperature estimated from the $B_p$ and $R_p$ spectra \citep{andrae23}
and integrate $P(T_{\rm eff})$ to compute the F-star probability in the same way as done for PS1.
We note that we do not use data from HSC-SSP for this purpose because bright stars for
flux calibration are often saturated in the HSC data.

These sky objects and flux standards are prepared by the observatory to ensure the quality
of the calibrations, and observers should use them by default. 
However, there are regions such as in the plane of the Milky Way where it is not easy to
find blank sky objects and sufficiently isolated, bright standard stars.
If a Principal Investigator (PI) targets such a region, we ask
the PI to prepare calibration objects before an observing run starts.

%----------------------------------------------------------
\subsection{Community Engagement}
\label{subsec:community_engagement}

While it is possible to build a new observing framework exclusively within the observatory,
it is more constructive and effective to do so by engaging with the community.
This is particularly the case for a versatile instrument like PFS that can be used for
a wide array of science goals; it is hard for the observatory to anticipate all use cases
a priori before the science operations start, and there are always users who
use an instrument in a way that is not originally intended.
Thus,  requests from the community on science operations are extremely important to incorporate 
in the new framework.

We had four separate discussion sessions at the annual Subaru Users’ 
Meetings in 2020-2024 to collect input from the community. We also held two PFS 
community meetings for further discussions, in particular on filler programs
(see Section\ref{subsec:fillers} for details of filler programs). In addition, we 
twice collected community input through online forms.
There were also offline iterations with groups of people interested in specific science cases.
These interactions with the community formed the 
basis of the new framework. The framework was further discussed with the Subaru Time Allocation Committee (TAC) and 
Science Advisory Committee (SAC) for further input. The PFS science operation framework  
detailed below is based on all these processes. This is the first time for Subaru to build a 
new science operation framework through such close communications with the community.

%----------------------------------------------------------
\subsection{Key Assumptions}
\label{subsec:key_assumptions}

In order to form a new observing framework, we start with two key assumptions.
One is that not all PIs will be able to use all the fibers efficiently at all pointings;
in other words, there will be unassigned, spare fibers due to lack of science targets.
This assumption has been motivated by
input from the community; there is a lot of interest in
follow-up observations of spatially sparse, rare objects selected from HSC-SSP imaging.
In observations of sparse targets, however,  a large fraction of fibers will be
left unassigned, leading to a low observing efficiency.
Such observations are still more efficient than using other traditional instruments
(e.g., targeting sparse targets one by one with a small-field instrument), but this is
certainly not the best way to use PFS. We thus have to come up with an idea to
assign more fibers to science targets in a given pointing and improve the efficiency.

The other assumption is that there will be a significant spatial overlap of targets between
different observing programs, as many of the targets are likely drawn from HSC-SSP.
In other words, there will likely be targets from multiple observing programs in a given PFS pointing.
This will especially be the case in popular fields such as Cosmic Evolution Survey (COSMOS; \cite{scoville07})
and Ultra Deep Survey (UDS; \cite{lawrence07}),
where ancillary multi-wavelength data are available and the fields can be covered by just
a few PFS pointings. Many of the survey-type programs can be drawn from the Wide layer of
HSC-SSP and we can expect significant spatial overlap there as well.

These two assumptions led us to a simple idea to execute multiple observing programs
in the same PFS exposure simultaneously, so that we can fill more fibers with science targets.
Different targets/programs may require different exposure times, but
they can be easily observed using multiple fiber configurations within the same pointing.
This idea, however, breaks Subaru's observing policy, in which only one program can be
executed in one exposure. A dedicated science operation framework for PFS thus became necessary.

%----------------------------------------------------------
\subsection{Framework}
\label{subsec:framework}

\subsubsection{Optimization metric}
Before we describe our science operation framework in detail, let us first define the observing
efficiency mentioned above because it is the metric we want to optimize. The exact metric
is actually rather complicated and can be found in He et al. (in prep), but to first order, it is
the sum of the numbers of fibers assigned to science targets weighted by the priorities. In other words, we want to
assign as many fibers as possible to scientifically important targets.
The actual optimization is done on a nightly basis rather than a visit-by-visit
basis using real targets for each night.

The target priorities are primarily set by the combination of program category and grade as summarized
in Table \ref{tab:categories}.
The category is chosen by the user and the grade is assigned by the TAC.
There are a few different categories for observing programs, such as normal program and filler program.
Normal programs go through the standard review process, while filler programs are evaluated
through a distributed review process among the PIs.
The selection of filler programs is more inclusive than the standard Subaru proposal selection mechanism,
and thus normal programs have a higher priority than filler programs.
Each program in each category is given a grade by the TAC, and higher-grade programs have higher priority.
Each observing program can also define internal, relative target priorities.
These relative priorities are considered only when there are multiple targets from the same
program for a given fiber (i.e., that fall within a given patrol region).

Again, our goal is to observe as many high priority targets as possible in each semester.
The optimization requires (1) an efficient algorithm to define PFS pointings to cover
high priority targets and (2) an efficient algorithm to assign fibers to targets.
We briefly mention these algorithms in the next Section, but the details are out of
the scope of the paper. The reader is referred to He et al (in prep).

\subsubsection{New Concepts}

Based on the assumptions laid out in Section~\ref{subsec:key_assumptions}, one obvious way to achieve a high observing
efficiency is to observe targets from multiple observing programs located in the same field
of view in the same exposure, given the likely overlap of target fields from different programs.
As mentioned earlier, this 'fiber share'\footnote{
This term has often been used in the discussions with the community, but it might be
slightly misleading because it is actually an exposure, not fibers, that observers share.
}
concept requires a fundamental change to Subaru's science operations, in which only one
program can be executed in one exposure.
Therefore, a new, efficient science operation framework for PFS is needed.

{\bf Observing mode:}
An immediate consequence of the fiber-sharing scheme is that the default
observing mode is queue. There is no other choice
because the classical mode just does not work; it is not practical for the observatory to
accept multiple observers for a given exposure either online or onsite
(and observers change from exposure to exposure). 
In queue, no observers have to attend and the observatory staff alone can carry out
the observation, making the overall operation significantly easier.
The observers need to submit all material needed for the observation in advance, and
we collect it as part of the proposal submission.

{\bf Observing time:}
One complication introduced by the fiber-sharing scheme is the definition of observing time
for each program.
The observing time is granted to accepted programs in units of nights or hours in the current framework. Because
we execute multiple programs and each program has its own requirements for observing conditions,
it is not clear how to charge one exposure time to each program (e.g., if 80\% of the fibers are used by program A
and 20\% by program B, should we equally charge them, or weight by their fiber fraction? What if only a fraction of
the targets meet the required observing conditions?).
In addition, each target may have a different total exposure time. An object can thus be observed multiple
times in different visits with different configurations of objects being observed, which makes
the time-charging more complicated.

To circumvent the problem, we introduce a new
unit for observing time: fiber hour (FH). A FH is an integration time for each fiber,
and if we expose one fiber for 1 hour, it is 1 FH. If we expose 2,400 fibers for
2 hours, the total is 4,800 FHs.
We will later introduce a concept of effective exposure, which accounts for observing conditions, 
but we ignore the observing conditions for simplicity for now.
The heart of FH is that it allows us to consider one exposure as a collection of FHs.
If we take a pair of 7.5min exposures, which is PFS’s default exposure time, then it is
$2400\rm ~fibers\times7.5~min\times2~shots / 60~min =600$ FHs.
Because one fiber is used for one target, a sum of FHs of all targets is the total
observing time for each program\footnote{
A target may be duplicated, i.e., multiple
programs may target the same object. In this case, we charge the FH to all of these
programs, although we have not implemented this policy in real operations yet.
If two (or more) programs target the same object for different total exposure
times, e.g., program A for 2 hours vs program B for 3 hours, we give the 3 hour data
to both PIs without charging an additional hour to program A.
}.

The concept of FHs significantly simplifies the accounting of the observing time
in the new framework. The TAC allocates FHs to each program, instead of nights or hours.
A program is considered done once the allocated FHs are executed.
The FH concept introduces, however, rather complicated issues during the TAC process.
We thus introduce another unit, Required Observing Time (ROT) only for the TAC purposes.
We will detail ROT in Section~\ref{sec:implementation_overview}.

{\bf Fixed individual exposure time:}
Another possible concern about executing multiple programs in an exposure is
that the individual exposure times are identical for all programs.
For example, if we take a 450sec exposure, all targets for all observing programs
in the exposure acquire 450sec of exposure
(we introduce effects of observing conditions in Section~\ref{subsec:effective_exposure}).
Some PIs may wish to take short individual exposures, while others may take long exposures.
The fixed individual exposure time is, however, unlikely a major problem because the total exposure time of a target is typically
longer than the individual exposure times, and we simply take multiple exposures to reach
the requested integration time. We may over-expose an object; e.g., if a PI requests 3,000s total
integration, we end up taking $450s \times 7 = 3150s$. The extra 150s is not charged to the program.
Also, completed targets may further be observed when there are no other targets to avoid
leaving the fiber blank. Such observing time is not charged either.

{\bf Classical observations:}
While queue is the default observing mode, there will be programs that cannot be executed
in queue. For instance, for a program targeting very bright sources, the individual exposure
time has to be shorter than 450s to avoid saturation. Also, some programs need to observe at fixed pointing centers.
We accept those programs in classical mode.
For classical programs, the observing time is allocated in half-night units like other instruments
at Subaru. We do not use FHs for classical programs. Also, classical programs can use their allocated
nights exclusively, and they do not necessarily have to share fibers with other programs, although
the observatory may still put filler targets on their spare fibers to improve the observing efficiency,
especially for time-critical observations (ToO), which we tentatively accept
in classical mode (Section~\ref{sec:implementation_overview}).
The individual exposure time does not have to be 450 sec and can be flexibly changed from
exposure to exposure. The new units and policies introduced for queue observing described
above are largely irrelevant for classical programs. We focus primarily on the queue mode in what follows.

\subsubsection{Filler Targets}
\label{subsec:fillers}
Even if we use the fiber-share scheme, we may not
always fill all the fibers at all pointings. For example, we can imagine a case where a target field of an accepted program
does not overlap with other programs (note that the TAC evaluation is based on science, not on
target fields). We need additional targets other than
those from accepted programs. These additional ‘filler’ targets are not just to fill spare fibers,
but they also offer a significant scientific
opportunity. In addition, not all observing programs can fit in the scope of normal queue
and classical modes (e.g., a program aiming to cover a very wide area), and
fillers can be a separate, useful category to accommodate such programs. In fact, fillers
were one of the major discussion items during the iterations with the community (Section
\ref{subsec:community_engagement}), demonstrating the potential of PFS for unique science cases.

We define two types of fillers: community filler and observatory filler. A community filler
program is a PI-led program. Unlike a normal program, a PI does not request observing
time. A PI instead just provides the observatory with targets to put spare fibers on.
A community filler program is suited to observe
targets that are spread over a wide area of the sky, for which a normal program cannot cover.
Note that a PFS normal queue program is limited up to 35 hours of observing time
including overheads\footnote{
A normal program in general is up to 5 nights. For a queue program, the weather factor
of 0.7 is folded in a priori, leading to a maximum request of 3.5 nights.
A night is considered 10 hours regardless of the season, and 3.5 nights translate into 35 hours
of observing time. The observing time here includes telescope and instrument overheads.
}.
It is also suited to observe sparse (low surface density) targets, which may otherwise be
difficult to justify using PFS.
A downside is that the spatial sampling of such targets will be  very inhomogeneous; as we discuss later,
filler targets do not drive the telescope pointings and they are observed only when they
happened to be located in the same field of view as high priority normal programs.
The sampling thus has to be statistically corrected for.
Another downside is that a PI cannot request a total exposure time; the exposure time
is driven by the accepted normal programs. Thus, one should not expect to get a longer
exposure than the default PFS exposure time of 15min (7.5 min $\times$ 2). In popular
deep fields such as COSMOS \citep{scoville07}, one may get a longer exposure than this, but it should not be taken for granted.
Due to the nature of the community filler category, we run community filler programs
for two semesters. In other words, we make a call for filler programs once a year as opposed
to the two calls per year for normal classical/queue programs. Table \ref{tab:categories}
summarizes differences between different categories.

Targets from community filler programs help us use more fibers than normal programs alone,
but the community filler targets are likely very heterogeneous in terms of both source
density and sky coverage because they are driven by specific science goals.
We still will not be able to use all available fibers all the time, and
we therefore need a separate target list that fully uniformly covers the observable sky from Maunakea with
a sufficient source density.

\begin{figure}
 \begin{center}
  \includegraphics[width=8cm]{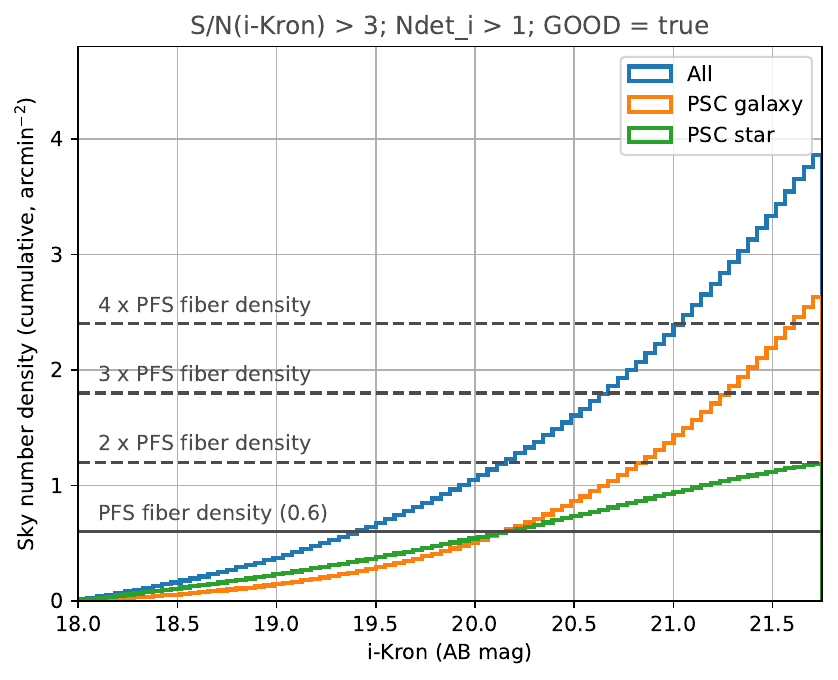} 
 \end{center}
 \caption{
   Cumulative surface density of objects from PS1 as a function of $i$-band Kron magnitude.
   The yellow and green curves are for extended and point sources, respectively \citep{tachibana18}.
   The blue curve is the sum of the two.
   The point source density is a strong function of the sky position, and the plot here is for
   the ELAIS-N1 field, which one of the HSC-SSP Deep fields, located at a Galactic latitude of about 45 degrees.
   We target all sources down to
   a certain magnitude limit irrespective of the extendedness.
   The horizontal lines show the (multiples of) PFS fiber density.
   At $i=21$ for instance,
   the source density is 4 times higher than the PFS fiber density, which is sufficient to
   serve as a filler sample.
   {Alt text: Cumulative surface density of sources per square arc minutes plotted against the i-band Kron magnitude.}
 }\label{fig:ps1}
\end{figure}

The observatory filler is introduced for this purpose.
We define a flux-limited sample of objects based on Kron magnitudes in the $i$-band
from the Pan-STARRS1 $3\pi$ survey DR2.
Fig. \ref{fig:ps1} shows the expected source counts.
Note that we do not distinguish point sources from extended sources
in the target selection; all sources down to
a certain magnitude cut are used as the observatory filler. The first PFS observing run
adopted a magnitude cut of $i=21.0$, but this may slightly change in the future runs.
There is also a bright magnitude cut of $i=18$ as of this writing to avoid contaminating
faint objects in the neighboring fibers due to fiber-fiber cross-talk and to reduce
scattered light inside the spectrograph.
We may relax this constraint with improvements in the data processing pipeline in the future.

%------------------------------

%-----------------------------------------------------
\begin{table*}[htbp]
  \begin{center}
    \begin{tabular}{cccccc}
      Category           & Competitive? & Proprietary period & Multi-semester & Priority & Drive pointing center?\\
      \hline
      Grade A            &  $\bigcirc$  & $\bigcirc$         & $\times$       & 1        & $\bigcirc$\\
      Grade B            &  $\bigcirc$  & $\bigcirc$         & $\times$       & 2        & $\bigcirc$\\
      Grade C            &  $\bigcirc$  & $\bigcirc$         & $\times$       & 3        & $\times$ (see caption)\\
      Community Filler   &  $\times$    & $\bigcirc$         & $\bigcirc$     & 4        & $\times$\\
      Observatory Filler &  ---         & $\times$           & $\bigcirc$     & 5        & $\times$\\
    \end{tabular}
  \end{center}
  \caption{
    Summary of proposal categories for queue programs. Classical programs are allocated time on
    specific nights and they can use the time exclusively, although the observatory may fill
    spare fibers with filler targets.
    The columns show the category, whether a proposal is
    selected through a competitive process, whether data have a proprietary period (1.5 years),
    whether a proposal runs over multiple semesters, priority order, and whether an accepted
    proposal can drive the telescope pointing, respectively. Grade A-C programs are called
    normal programs and are selected through a competitive process. Grade A-B are accepted programs
    and they determine PFS pointings. The grade C programs are unaccepted programs,
    but they are kept in the queue and they may be observed when there are spare fibers.
    Also, they may drive the telescope pointing when there is nothing to observe from the grade A and B programs.
    The community filler is selected
    through a distributed review process. It is a rather inclusive process to make sure that the observatory has
    a large number of targets to fill spare fibers.
  }
  \label{tab:categories}
\end{table*}

%------------------------------
\begin{figure*}
 \begin{center}
  \includegraphics[width=18cm]{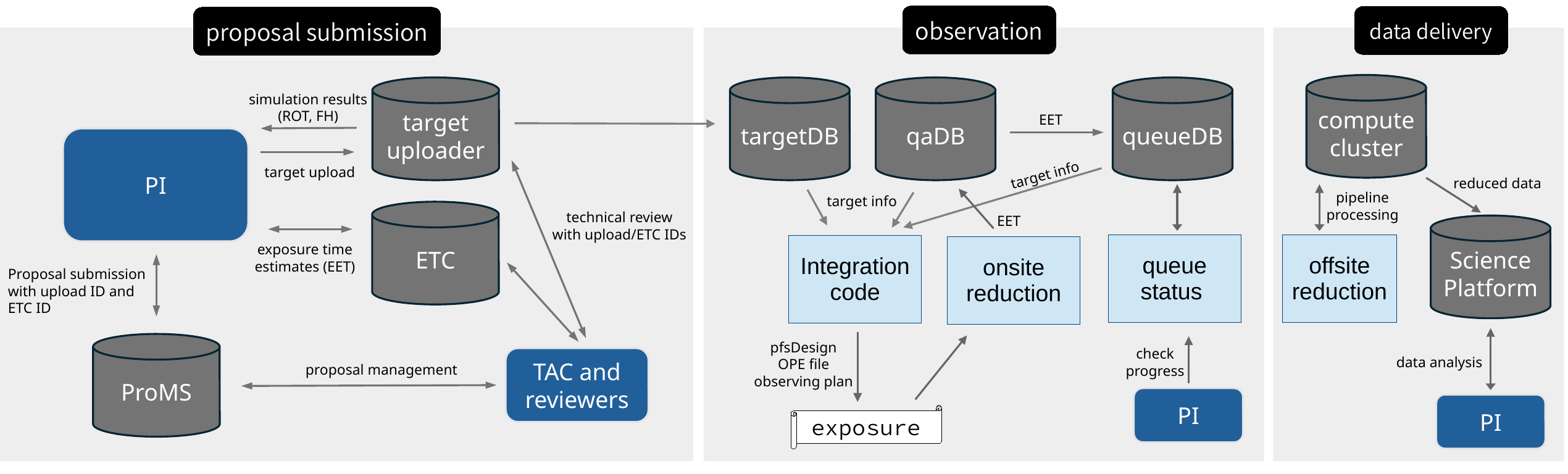} 
 \end{center}
 \caption{
   Schematic view of the workflow from the  proposal processes to data delivery. From left to right, the panels show phase 1 (proposal submission phase),
   observation, and data delivery phase. See the main text for details of the workflow.
   {Alt text: A schematic drawing of the workflow.}
 }\label{fig:flow}
\end{figure*}

 %----------------------------------------------------------
\subsubsection{Effective Exposure Time}
\label{subsec:effective_exposure}

In queue mode, the effectiveness of an exposure has often been a binary classification; if an exposure
is taken under (or better than) requested conditions, the exposure is marked good. Otherwise, bad.
This binary classification is simple but is not ideal for PFS because different observing programs
may require different conditions. For instance, program A requires excellent conditions, while program B
can be executed under poor conditions. If we execute both programs in the same exposure under
average conditions, the exposure is considered good only for program B. However, the exposure
is not useless for program A, and targets from program A do gain a signal-to-noise ratio (S/N).
If an exposure is taken under excellent conditions, program B gets a much better signal than
requested and may be completed with fewer exposures. The binary classification cannot capture
these important details.

We choose to use the S/N to define the effectiveness of an exposure so that we can account
for different observing conditions in different exposures. As PFS is a fiber spectrograph,
the signal can be affected not just by the sky transparency, but seeing as well.
E.g., if the seeing is worse, less light goes into the fiber. Guiding error
and fiber positioning error also reduce the signal. All these effects are degenerate,
and we define the signal as the total throughput as measured in the data.
We use flux standards for the throughput measurement in each exposure.
Point sources like flux standards give a pessimistic and thus conservative throughput estimate;
the reduction in S/N due to poor seeing is smaller for extended sources than for point sources.
It is, however, difficult to deal with throughput for an arbitrary size and profile of
an object for science operations, we choose to adopt the simple assumption of point sources here\footnote{
DESI uses object profiles to aperture-correct spectroscopic fluxes to a fiducial set of aperture and
seeing sizes \citep{guy23}.  As a general purpose instrument, PFS does not do the same because
not all users have structural information of their targets with sufficient accuracy.
}. For the noise measurement, we simply measure the average noise level from the sky objects
from each exposure.

As S/N is wavelength dependent, we define 
a small wavelength window in each arm to measure the S/N.
The windows are chosen to avoid regions where the sky lines and telluric absorption are too strong.
To be specific, we use $525-555$~nm for the blue arm, $835-860$~nm for red (including medium resolution),
and $1040-1070$~nm for the near-infrared arm.
The user is asked to choose the 'reference arm' for each of their targets based on the science goals.
The reference arm here can be either one of blue, red, medium-resolution red, and near-infrared.
We note that effects of the Moon are included in the S/N; if the Moon is up,
the background level increases, and accordingly the S/N decreases, the amount of which is
dependent on wavelength and thus reference arm.

In order to translate the measured S/N into exposure time, we define the fiducial throughput
and noise level. They are measured for each arm from engineering observations executed under
typical observing conditions. For reference, we find that the fiducial throughput and noise
are consistent with estimates from the exposure time calculator (ETC)
for Moon phase\footnote{Here, the new Moon is phase=0, and full Moon is phase=1.}= 0.15, Moon
distance = 60~deg, and seeing=0.8 arcsec.
If an exposure is taken under the fiducial conditions, EET is equivalent to the exposure time.
If conditions are worse, EET is less, and vice versa.

As a more specific example, consider a 900~sec exposure. If the measured throughput is 80\% of the fiducial throughput,
then the collected signal is 80\% of the fiducial.
If the noise level is 10\% higher than
the fiducial level, the relative S/N is $0.8/1.1=0.73$ compared to a hypothetical exposure taken
under the fiducial conditions. As S/N scales with square root of the exposure time in background-limited observations,
we consider the effective exposure of $900 \times 0.73^2 = 480$ seconds.
We charge this 480 seconds to the user.

We ask the user to estimate the exposure time for
an object using an (online) ETC assuming the fiducial throughput and noise level
in the proposal submission phase (Section~\ref{sec:proposal_submission_phase})
because an exposure time under the fiducial conditions is the EET.
As we discuss in detail in the next Section, we do not let the user define required
observing conditions for each target. The differences in the observing conditions are
effectively absorbed into EET.
For objects that require excellent conditions, the user
requests a long integration time. On the other hand, for objects for poor conditions, the
user requests a short integration.

%--------------------------------------------------------------------------------------------------------
\section{Implementation Overview}
\label{sec:implementation_overview}

Following the concepts introduced in the last Section, we now describe our implementation
in the order of the actual workflow.
Fig.~\ref{fig:flow} is an overview chart of our processes/tasks. Also, Table \ref{tab:categories} gives
a comprehensive summary of the proposal categories and their priorities for queue programs discussed in the previous Section.
We emphasize that, as we mentioned earlier, the science operation framework will evolve
in the future and the reader is referred to the PFS instrument page as well as call for
proposals at the Subaru website for the latest information.
Note as well that that we focus on PFS specific policies and operations,
and we do not go into details of Subaru-generic operations.

%----------------------------------------------------------
\subsection{Proposal Submission Phase}
\label{sec:proposal_submission_phase}

All Subaru proposals are submitted through the Subaru Telescope Proposal Management System (ProMS).
PFS uses the same system with small modifications in the submission form to accommodate a number
of changes and additional information required for the subsequent TAC process
and science operations such as {\tt uploadId} (see below).
The primary category of PFS observations is, like the other instruments at Subaru,
the normal program, which is selected through a competitive process; a PI submits a proposal, the TAC reviews
and grades it, and then the observing program is allocated time.
Another category is filler, which goes through a different process. There are a few key tools needed
for a PI to prepare a PFS proposal. In this Section, we start with these tools and then move on
to discuss the TAC and actual observing processes.

%----------------------------------------------------------
\subsubsection{Exposure time calculator}

One key tool for a PI to prepare a PFS proposal is the ETC.
It allows the PI to estimate the total exposure time required to
achieve their goals. There are several ingredients for ETC, such as system throughput,
models for the telluric absorption and night-sky lines, and so on.
Our fiducial telluric model incorporates the continuum component from the Kitt Peak model
\citep{hinkle03} and the line component from the Gemini Maunakea model for PWV = 3 mm \citep{lord92}.
Our fiducial sky brightness model includes both continuum and
emission line components. The continuum component is based on SDSS/BOSS spectra fitted to
an analytic functional form (Jim Gunn; Private Communication). The emission line component is
based on the UVES atlas \citep{hanuschik03} at $\lambda < 1040$~nm and a theoretical model
\citep{rousselot00} at longer wavelengths. The model is calibrated to reproduce the observed
broadband sky brightness fluxes. For the effect of moonlight, we employ a two-component
(Rayleigh + Mie) moonlight scattering model based on \citet{krisciunas91}.

There are other parameters that can be adjusted in the ETC such as airmass, seeing, transparency, moon phase,
and moon distance. However, we ask the PI to fix all of them to the fiducial values set
by ETC. To be specific, we adopt seeing FWHM=0.8 arcsec, transparency=0.9,
moon phase=0.15, moon distance=60~deg, and target elevation of 55~deg.
This is because, unlike typical queue programs at other telescopes,
we do not ask the PI to define observing conditions for the targets.
As discussed in Section \ref{subsec:effective_exposure}, we use EET to evaluate the effectiveness of an exposure, which
folds in the effects of observing conditions by construction, and
%If a different PI  adopts
%a different combination of observing conditions in the ETC, it is very hard to interpret the required
%exposure time.
we thus fix the parameters for all programs. It is important also from the time allocation perspective
that we have only one definition of observing time.
The throughput and background levels in each arm are estimated from engineering observations executed
under reasonable conditions as mentioned in Section~\ref{subsec:effective_exposure}.
In other words, the ETC is calibrated with real observations, ensuring  that an exposure time estimated
using ETC is equivalent to an EET.

There are a number of pre-defined spectral templates for galaxies and stars, which can be
normalized by specifying, e.g., a flux density at a given wavelength (and redshift for galaxies).
The PI can also upload an arbitrary spectrum, which in many cases is an expected spectrum of
the target. In addition, detections for emission lines can be evaluated.
In this way, the PI can estimate the exposure time required to achieve the science goal.

\begin{figure*}
 \begin{center}
  \includegraphics[width=18cm]{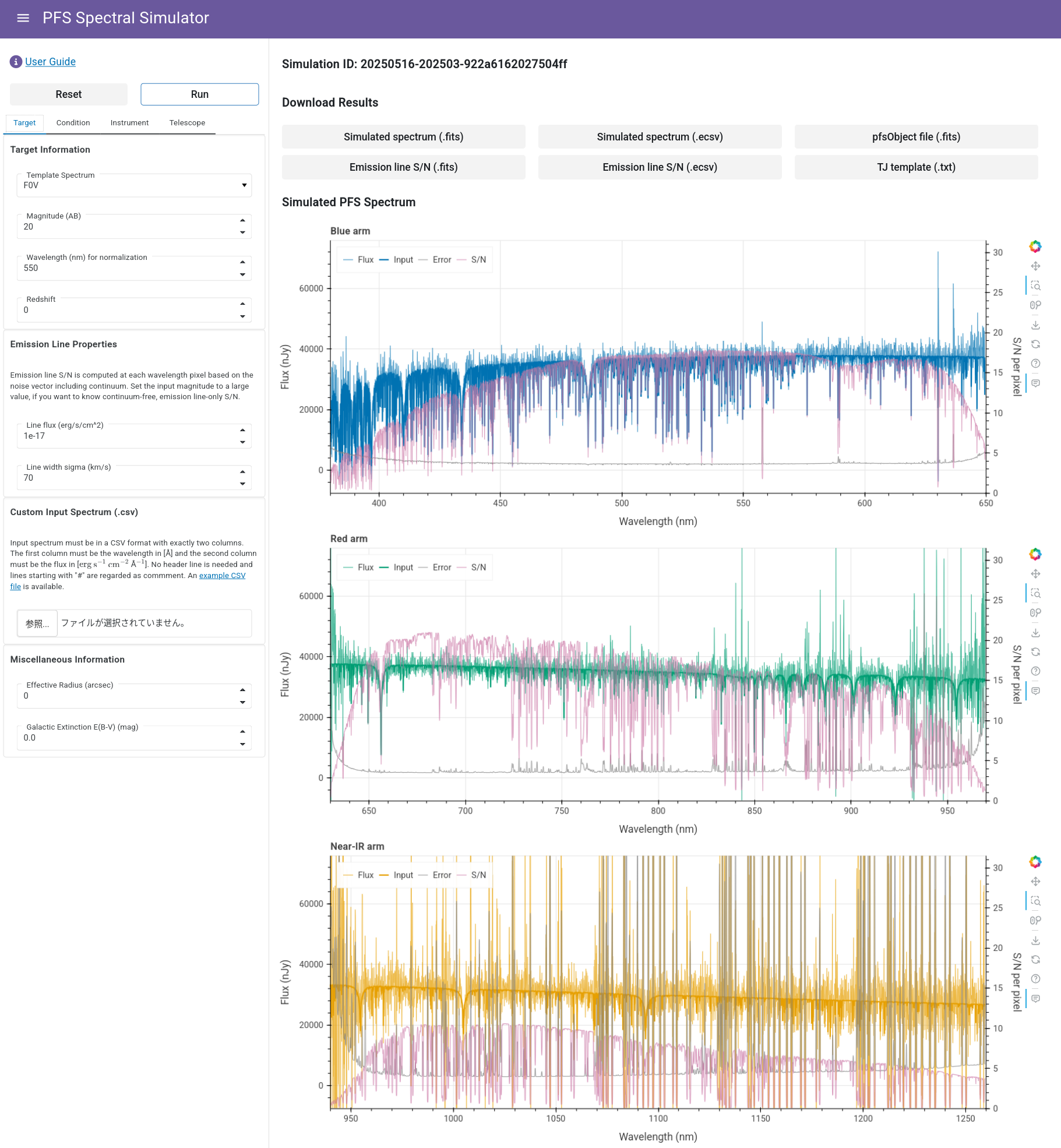} 
 \end{center}
 \caption{
   Screenshot of the online ETC with an example exposure of 1 hour for a F0V star normalized to 20th magnitude
   at 550nm. The default observing conditions are assumed (see the main text for details).
   The panels and tabs on the left show
   the ETC configurations and the right panels show the expected spectra for each arm.
   In each of the right panel, the input template spectrum, object spectrum (which is the input spectrum
   perturbed by a realization of the noise),
   noise spectrum, and signal-to-noise ratio are shown as input, flux, error, and S/N, respectively.
   Although not shown in this figure, ETC also generates expected S/N of an emission line as
   a function of wavelength. The user manual is available from the link on the top-left.
   {Alt text: A screenshot of a working example of the ETC.}
 }\label{fig:etc}
\end{figure*}

There are two versions of the ETC: one offline and one online. The functions are
nearly identical, but the PI is asked to use the online version,
which issues an ID for a given ETC run, for proposal submission. This allows
the observatory to check the technical feasibility during the proposal review process.
The offline version can be used for, e.g., intensive tests against a large number of templates.
A screenshot of the online version is shown in Fig. \ref{fig:etc}.

%------------------------------

%----------------------------------------------------------
\subsubsection{Target Uploader}
Another key tool is the target uploader. Once the PI has a target list with total
exposure times estimated using the ETC, the PI has to run a simulation of the proposed
observation to estimate the total observing time including overheads. The
online target uploader accepts a target list from the PI, runs a survey
simulation, and submits the target list to the observatory. Recall that the PI
does not need to include calibration objects such as sky objects and standard stars,
as the observatory prepares them (see Section~\ref{sec:instrument}).

The survey simulator, which is a major component of the uploader, simulates the observation
and estimates the total observing time. It identifies reasonably optimal pointing centers to cover
the uploaded targets. At this point, we do not run the full optimization because
(1) it is time-consuming, and (2) it is not very meaningful to do it for only one program
(the survey optimization makes most sense when we have full target lists from all
accepted programs). We identify the pointing centers by applying a kernel density estimate (KDE)
to the targets weighted by internal priorities.
The kernel size is set to the field of view of PFS.
We ignore the fact that the PFS field of view is hexagonal and adopt a circular Gaussian kernel.
We then use the most significant peak as the PFS pointing center, assign fibers to
the targets (the fibers are assigned to randomly chosen targets within
the field of view), observe, remove targets that reach the required total integration,
apply KDE to the remaining targets, and repeat. 
Due to the random fiber assignments, the results slightly differ from run to run,
but the difference is sufficiently small in most cases that other factors such as weather
conditions can easily bring in larger fluctuations in the real observing time.

\begin{figure*}
 \begin{center}
  \includegraphics[width=18cm]{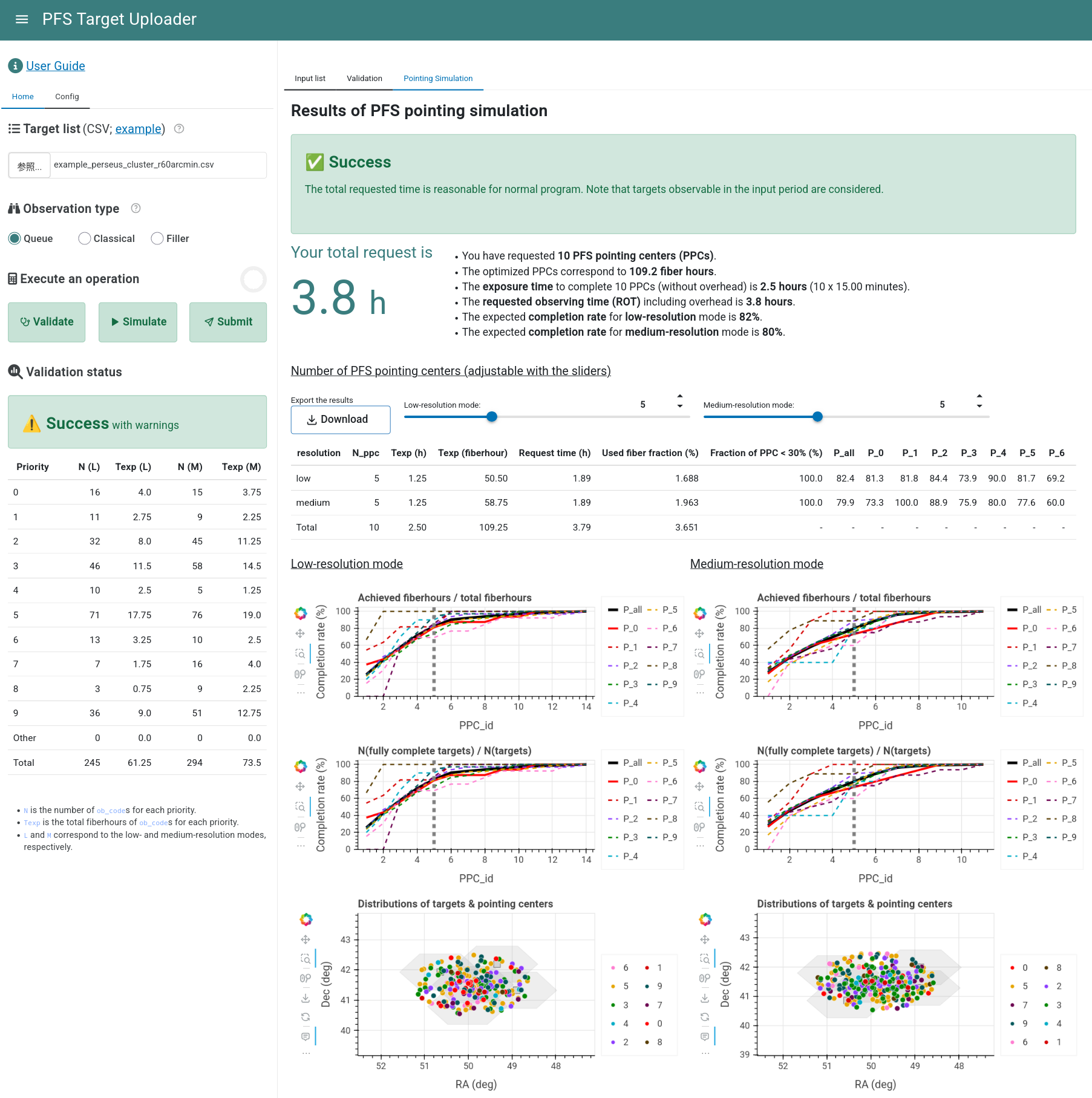} 
 \end{center}
 \caption{
   A screenshot of the online uploader. If the user uploads a target list,
   the uploader validates it and runs a simulation of the observation.
   The target information is shown on the left and the simulation result is
   summarized on the right. The user can move the sliders near the center of the page to change the number
   of pointings for low and medium resolution modes separately, and the total observing time
   changes accordingly. In this example, ROT is 3.8 hours.
   {Alt text: A screenshot of a working example of the online uploader.}
 }\label{fig:uploader}
\end{figure*}

The simulator generates several useful plots and metrics to evaluate the efficiency
of the proposed observation. See Fig. \ref{fig:uploader} for an example.
The simulator shows the completeness as a function of the number of exposures, and
in most cases, the target completeness, which is a fraction of fully observed targets
among all targets, increases rapidly in the first $\sim50\%$ of the exposures
(the exact number depends on the target distribution) of the required pointings and then
the progress slows down. This is because the fibers are efficiently assigned to
targets in the early phase when there are a lot of targets, but in the later phase,
there remain only targets that are not observed due to collision with nearby targets.
The collision here means that multiple targets are located in the patrol area of a single fiber,
and that single fiber has to observe them one by one.
In the late phase, most fibers are left unassigned and only a small
fraction of the fibers are used to observe the objects left from the early phase due to collisions.

The PI can choose how many exposures to take. Many PIs take exposures to
achieve a reasonably high completeness while keeping the number of exposures
($\sim$ total observing time) relatively small. The example case in Fig. \ref{fig:uploader}
is the case where the target
completeness is about 80\%. The total number of exposures is 10, which is
only 40\% of the exposures required to complete all targets.
The number of exposures chosen here linearly correlates with the observing time required for the program.
The simulator gives the total observing time including overheads in two ways;
total FH and ROT. We will explain why we need both of them in the next subsection.

The last operation for the PI is to submit the target list. Unlike 
Subaru proposals for other instruments, PFS requires the final version of the target list together
with the proposal simply because the list is needed for the time allocation
(but see an exception below).
The observatory does not allow the PI to update the list after submission
unless there is a very good reason to do so, because a change in the list
means a change in the observing time. Accordingly, there is no 'phase 2' for PFS.
Once the list is submitted, the uploader issues an ID ({\tt uploadId}). The PI
must include this ID in the proposal.

An exception can be made to observing programs that do not
have a target list at the time of the proposal submission such as normal proposals
to follow up transient sources (ToO proposals are a separate category and we will
discuss their implementation later in this Section).
In such a case, we currently ask the user to submit a classical
proposal instead of queue and send the target list a month before the assigned
observing run. This is not a very flexible scheme as the deadline for target
submission ideally has to be much closer to the observation for such programs.
This is one of the points that the observatory plans to improve in the future.

The same uploader is used for community filler programs. The simulator does not run
for fillers because a filler program does not request an observing time.
The uploader simply checks the format of the uploaded target list and
issues an {\tt uploadId}.
Normal programs are called for every half a year, but community filler
programs may target a large area of the sky, which cannot be covered in
a single semester. Also, we want to reduce the workload of the
TAC for the community filler review process. For these reasons, we make
a call for filler targets once every year. In other words, community filler programs
run for two semesters.

All of the uploaded targets are stored in a database called {\tt targetDB}.
{\tt targetDB} includes not just science targets but calibration objects and
guide stars as well. Each observing program is assigned a unique catalog ID,
which is used in the data processing. We query the {\tt targetDB} when
we make plans for the night and make fiber configuration designs
(see Section \ref{subsec:during_a_run}).

%------------------------------

%----------------------------------------------------------
\subsection{Time Allocation}

After the submission deadline, TAC will evaluate and grade the proposals.
The PFS’s fiber-share policy introduces a new difficulty in the time allocation; due to the statistical
nature of the fiber-share scheme, we do not know a priori how many nights we actually need to
complete a program.
If, for instance, two programs with similar exposure time per object target the same field,
we expect a high observing efficiency, and  the total observing time to complete the two programs will be relatively short.
On the other hand, if their fields do not overlap, we will need more time to complete them
because we need to execute them separately.
In other words, the actual observing time is dependent on which programs are accepted.
This is a problem because we need to estimate the total observing time of each program
for time allocation purposes, e.g., program A requires 2 nights, B requires 1.5 nights, etc,
so that we can fill the available nights with an appropriate number of programs, 
but the numbers depend on which programs to accept.
We are in a chicken-and-egg situation.

One approach to make a rough guess for the actual observing time
is to make multiple possible combinations of accepted programs, run the survey
simulation, and estimate the observing time during the time allocation process.
However, the time available for the time allocation process is rather limited
and it is not practical, at least on a short term, to carry out many simulations.
We choose to introduce a simple way to define the actual observing time,
which is to assume classical mode without sharing fibers with other programs
and adopt the requested time as ROT.
The target uploader gives ROT because it assumes no fiber share with other programs,
e.g., in the case shown in  Fig.~\ref{fig:uploader}, it is 3.8 hours.

ROT will be an over-estimate of the actual observing time because it ignores
the fiber-share scheme. However, it offers
a simple and fair way to define the observing time. Such a simple definition is particularly important when
we consider time-exchange programs; Subaru, Gemini, and Keck telescopes exchange
time in each semester, and there has to be a reasonable, homogeneous way to define observing time
regardless of instrument and observing mode across the observatories. Time exchange
between classical programs are simple, but fiber-share queue mode is not simple
at all. In addition, the fact the proposal deadline for each community is different makes
the chicken-and-egg problem even more difficult. Given all of this, we choose to adopt the classical
observing time as ROT for all PFS programs at least as a short
term solution.

We note that ROT is used only for the time allocation purposes.
As mentioned earlier, ROT is likely an over-estimate, and FH is a better measure of
real PFS observing time. TAC and observatory thus do the time allocation work using
ROT, and grants actual observing time for each PI in FH.
We may be able to introduce an effective scaling
factor to match (at least statistically) ROT and FH.  If we can establish
such a mapping in the future, we may be able to partially solve the chicken-and-egg problem.

For community filler programs, the evaluation process is different. As there may
be many filler proposals, we introduce a distributed review system, in which PIs
review each other's proposals, thus reducing the workload of the TAC.
This is the first time for Subaru to use the distributed review system.
There is a risk of being reviewed by non-experts, but
the community filler evaluation process is, unlike normal programs, a rather
inclusive process because the more filler programs PFS has, the better.
In addition, the review process is good experience for junior researchers.

During the time allocation process, we need to make sure that there are sufficient
targets to cover the R.A. range visible in the semester. PFS nights are allocated by
careful considerations of instruments (recall Subaru is a general-purpose telescope
and we often change instruments), program priority, target visibility, moon phase, etc.
This is a rather complicated process, and the PFS proposals were selected conservatively
in the first semesters just to be on the safe side. As a result, we faced a situation where
we did not have accepted programs to drive the telescope pointing in a small time window,
and we had to use grade C programs (see Table~\ref{tab:categories}).
In order to avoid such a situation in the future, we plan to run a semester-wide simulation
during the TAC process to make sure that we have sufficient targets throughout the semester.

%----------------------------------------------------------
\subsection{Science Operations}
\label{sec:post_submission_phase}

After the TAC process and the results have been passed on to the PIs, there is nothing further
for the PIs to do.
In other words, there is no phase 2 because the observatory already has
the final version of their target lists from the proposal submission phase (phase 1).
If the PIs have a good reason to make a small change (e.g., a reduced amount of observing time is allocated),
such a change can be made. But major changes such as adding new targets are not allowed at this point.

After the target lists are finalized and PFS observing runs are allocated, we start to
prepare for real observations. This subsection describes our first implementation of
the PFS science operations in the order of actual workflow. Once again, refer to the PFS
website for the latest information because the actual implementation is the place where
improvements are likely to occur most frequently.

%----------------------------------------------------------
\subsubsection{Before a run}
\label{subsec:before_a_run}

The science operation procedures are different for different program categories.
For normal queue programs, we carry out a run-wide simulation using all the targets
from accepted normal queue programs observable in the run to make a rough plan for
the run. To be specific, we first check for over/under-abundance of targets.
As mentioned earlier and summarized in Table \ref{tab:categories},
only grade A and B targets contribute to determine the telescope pointings.
Grade C targets contribute only if there is nothing else to observe
from grade A and B programs. Such a situation may happen in a specific time window
during a run, and we want to know the time window a priori so that we can include
grade C targets in the pointing determination there.
Another aim of the run-wide simulation is to predict completion rates so that
we can consider the observing strategy in advance.  Finally, the simulation
can be used as a back-up observing plan in case of any problems in the daytime work
(e.g., if daily preparation for the night such as designing fiber configurations does not
finish in time due to an unexpected power outage).

PIs have to submit their final target lists in phase 1 even for classical programs.
However, if a program cannot prepare a target list for scientific reasons
(e.g., transient targets selected from on-going imaging surveys), we collect targets from the PI a month before the run.
For all classical observations, the observatory is responsible for making fiber designs
because calibration objects as well as guide stars have to be selected in each exposure.
We do not let the observers choose them for a stable operation of the instrument as well as for
ensuring good calibrations by the pipeline.
We make designs and iterate with the PI to make sure that the proposed
science goals can be achieved.

For filler programs, there is nothing for PIs to do in preparation for the run. The observatory assigns spare
fibers to filler targets whenever available in queue observations. The filler
programs may also be included in classical observations when spare fibers are available.

%---------------------------------------------------------
\subsubsection{During a run}
\label{subsec:during_a_run}

There are both daytime tasks and night-time tasks during a PFS observing run.
A PFS run is typically 2 weeks centered on the new Moon and is scheduled by the observatory
as part of the TAC process. On the starting day of the run, the Prime Focus Instrument,
which houses the fiber positioners, guide cameras, as well as calibration lamp system,
is installed on the telescope during the day, making the whole instrument ready for
night operations.

We make a nightly plan during the daytime. We first compute EETs from the previous night
(see below for the night operations) and update the total EETs for each target and for
each program. This is managed in the quality assurance database ({\tt qaDB}).
The {\tt qaDB} records the quality of each exposure measured from the 2d pipeline,
which reduces raw 2d data to fully calibrated 1d spectra
(recall that, for EET, we measure the background noise from the sky fibers and throughput
from the calibration stars and compare them with the fiducial numbers in each reference wavelength window).
In addition to the QA results, seeing and transparency measurements from the guide cameras are also stored,
which can also be used to check
the quality of an exposure.

After the {\tt qaDB} is updated, we simply sum up the EETs across each exposure to see the progress of each target.
For objects that reach the required EET, we consider the objects done. Similarly,
once we reach the allocated FH of a program, we consider the program complete.
The progress of each target/program is monitored in the queue database ({\tt queueDB}).
There is a web user interface called the queue status page, where PIs/co-Is of queue programs
can login and see the EET of each target thus far as well as
the overall program status. The observatory staff can see the status of all the programs and
we use that information to tweak each night's observations if needed.

After the target/program statuses are updated, we run a piece of software called ``integration code'' to define
pointing coordinates, assign fibers, determine the schedule, and generate fiber configuration design files.
During this process, the target visibility and other instrumental constraints are accounted
for, and an OPE file, which is a collection of instrument and telescope commands for science operations,
is also generated together with an observing plan.
These output files are checked by the daytime staff and are uploaded to a summit computer
for operations at the summit facility before the evening.
The night staff executes the fiber designs prepared during the day. In order to reduce
the workload of the night-time staff, we do not flexibly change the observing plan depending
on the conditions. We instead simply follow the plan from the daytime staff. There is obvious
room for future improvements here, but a lot of development work would be needed to flexibly change
the plan and make new fiber configuration designs on the fly.

For classical nights, we already have fiber configuration design files that the observers
have validated beforehand (Section \ref{subsec:before_a_run}), and the observer has an observing plan.
The night staff follows the plan prepared by the observer. As in the queue mode, for the time being,
we do not generate new designs during the night to reduce the workload of the night staff.

During the observation, a real-time data reduction process runs soon after an exposure is taken.
The data are first registered to a local repository on a server in the Hilo base facility and
then the processing starts. Details of the pipeline will be discussed elsewhere,
but the pipeline first removes instrumental signatures such as bias and dark and then applies
a small adjustment to the fiber traces and wavelengths from the expected positions.
It then extracts 1d spectra from the calibrated 2d images.
The sky subtraction is performed using the sky fibers (Section~\ref{subsubsec:sky_objects}) and
the three arms are combined to make single, continuous spectra of objects.
The pipeline proceeds with flux calibration using the standard stars (Section~\ref{subsubsec:flux_standards})
and generates fully calibrated 1d spectra for all objects in a given exposure.
After this 2d processing is done, we carry out onsite quality assurance tests.
An important task there is to compute the EET for each visit
as discussed earlier. The EETs are then used for daytime planning for the next night.
We repeat these daytime and night-time tasks described here to operate the instrument during a run.
The onsite observing log system ({\tt obslog}) provides the observers and observatory staff with useful capabilities
to check targets and data quality. One can leave comments to each exposure that can be useful
for post-observation processing.

During a run, a ToO trigger may occur. A ToO program is normally for a small number of
objects and we may give most of the fibers to filler targets, but we accept a ToO trigger
whenever PFS is on the telescope. In the current operation rule, a PI has
to make a ToO trigger by 9am (HST) on the day before, so that
the observatory has time to iterate with the PI, make fiber configuration designs,
and validate them. In the first PFS run in 2025 Mar-Apr, the first ToO program
was triggered \citep{zhang25} to search for an optical counterpart of a gravitational wave signal.
While the ToO observation itself was successful, we saw a lot of room for improvements.
For instance, a ToO program is awarded time in units of half-nights. However,
not all ToO programs require a half night, and we hope to be more flexible and
grant time in units of hours, which will not be too difficult given that
the default observing mode of PFS is queue.

%----------------------------------------------------------
\section{Data Distribution}
\label{sec:data_distribution}

After a PFS run is over, all data are archived. Due to the complexity of the PFS
science operations, PFS data archiving is also difficult and complicated.
Also, for the first time for a Subaru instrument, the observatory reduces the data
from each observing run for the users. This Section summarizes these post-run activities.
To be specific, we discuss the raw data archiving (Section \ref{subsec:stars}),
user support (Section \ref{subsec:pfs_help_desk}), and delivery of reduced data to
the users (Section \ref{subsec:pfs_science_platform}).

\subsection{STARS}
\label{subsec:stars}

All Subaru data are archived. STARS (Subaru Telescope ARchive System;  \cite{winegar08}) is the primary archive
of raw data from all instruments. All science data other than calibrations have a proprietary period of 1.5 years, although there are exceptions.
For Subaru observations with other instruments, where one program is executed at a time,
this proprietary period policy is simple and easy to operate. However, in the fiber-share scheme of PFS,
a single exposure is shared by multiple PIs. This makes the data distribution complicated.

A PFS exposure includes a total of about 2400 fiber traces in 4 spectrographs with 3 arms each.
Some of the fiber traces are calibration spectra, but the others are
science spectra for multiple programs. If we deliver a raw image file to all the PIs,
they can process fibers used for other programs. This breaks the idea of
the proprietary period and makes it easy to scoop other (potentially competing) programs.
We want a PI to see spectra of their own targets only.

Information about fiber configuration (e.g., which fiber is for which target from
which program) for each exposure is stored in a file called {\tt pfsConfig}. It also
includes additional information about telescope pointing, target/measured fiber positions, fiber
status, etc. It is the central source of fiber configuration information for data processing.
A way to mitigate the problem mentioned above is to deliver a custom {\tt pfsConfig} file
for each PI with information of targets from other programs masked. This way,
as long as a PI uses the PFS data reduction pipeline, only spectra for the PI’s
targets are processed. We consider it the most reasonable solution and adopt
it in our operations.

It is of course possible to reduce PFS data without using the standard pipeline. We 
expect that the vast majority of people will not do so, but it is technically possible to
manually reduce spectra taken for others. To eliminate this small possibility, we ask
all PIs not to do this sort of manual processing. We request that all PIs who submit a PFS proposal
agree with this policy.

We are now in a position to discuss the STARS archiving system. As mentioned above,
one science exposure belongs to one observing program for all past/current Subaru instruments.
The STARS system thus has assumed that one proposal ID can be assigned to one exposure at
the database/infrastructure level. However, PFS breaks this rule, and a major revision was
needed to the system. After a major upgrade, the system now accepts multiple programs for a given science exposure,
and the PIs of targets observed in that exposure can retrieve the raw image data from STARS
in the same way as for the other instruments.
The reader is referred to the STARS website\footnote{\url{https://stars2.naoj.hawaii.edu/}}
for details.

Following the Subaru's archive policy,
raw data go public after 1.5 years of proprietary period, and anyone is welcome to use them.
The original {\tt pfsConfig} files with full target information will be released at the same time.
We recall that the observatory filler does not have a proprietary period (Table \ref{tab:categories}).
To eliminate the risk of open-use programs that are observed in the same exposures as observatory filler getting scooped,
we do not make the observatory filler raw data available at STARS during the standard proprietary
period. Instead, the reduced, science-ready data are made available through the science platform
after each observing run (Section \ref{subsec:pfs_science_platform}). The community filler programs
are the same; no raw data can be retrieved from STARS and only the reduced data are available.

%----------------------------------------------------------
\subsection{PFS Help Desk}
\label{subsec:pfs_help_desk}

While we expect the vast majority of the users will use the reduced data from the PFS Science Platform (SP),
we leave room for pipeline processing by the users. We have made an online pipeline tutorial\footnote{
\url{https://subaru-pfs.github.io/pfs_helpdesk_tutorial/}
}
to walk them through the pipeline from installation to data analyses. The pipeline is still
in an active phase of development and we expect frequent updates to the tutorial.
We prepare a help desk to further support the users by emails.
The pipeline is open-source software\footnote{\url{https://github.com/Subaru-PFS/}}
and anyone can use it. However, building calibrations requires a good understanding
of the instrument and the code to build calibrations is rapidly evolving.
The pipeline is not very easy to use in practice at this point, unfortunately, but
we hope to make it easy to run by anyone in the future. In any case, the PFS help desk is
there to help users.

The help desk also offers support to the PFS SP described below. We encourage the users to
use the pipeline functions to analyze the data and thus the data analyses are closely tied to
the pipeline. It is therefore efficient for the same help desk to support both the SP users
and the pipeline processing. The help desk also deals with questions about the functionalities of the SP itself
such as login and ssh connections. Some of the frequently asked questions are summarized in
the getting-started documents.

%----------------------------------------------------------
\subsection{PFS Science Platform}
\label{subsec:pfs_science_platform}

The PFS data processing is more involved than any of the Subaru instruments due to
the complexity of the instrument. In particular, it is still a challenge
to build all calibrations such as {\tt fiberProfile} (which describes the spatial profile of
each fiber as a function of wavelength used for extraction), especially in the near-infrared
where active development work is still on-going.
The observatory thus releases the reduced data to the users.
This is important from the perspective of the proprietary period; if the observatory
delivers the processed data, there will be many fewer people processing their data by
themselves, reducing the risk discussed in the previous section. The most important point,
however, is to allow PIs to kick-start their science soon after the observing run.
This is the first time that Subaru has released fully reduced data.

After each observing run, we first build calibrations needed to
process science data. After validating the calibrations, we move on to
process science data. This per-run processing is intended as a preview release,
meaning that its data quality is on a best-effort basis, although we provide
the user with quality assurance test results.
We also intend to use the per-run release as a pipeline development cycle;
we process and deliver the data, the observers look at the spectra, identify problems,
report back to us, and we fix the problems in the next release. This is a $\sim2$ month
cycle as of this writing, which may be too short for fixing big problems, but it is long
enough to fix small ones.

There is also a semester-wide processing run, which is intended
as a more formal release including all visits taken in that semester. Some of
the observing programs run over multiple semesters (e.g., community filler), and there will
be a multi-semester processing campaign, although details are yet to be defined as of this writing.

In addition to the 2d processing, we also run 1d pipelines. There are two 1d pipelines;
the redshifting pipeline developed by the team at Laboratoire d'Astrophysique de Marseille
(a.k.a. LAM 1d pipeline), and the stellar pipeline developed by the SSP Galactic Archaeology
team (a.k.a. GA 1d pipeline). The LAM pipeline \citep{schmitt19} measures emission
and absorption lines in addition to redshifts, and
it also classifies objects into galaxies, stars, and quasars. The GA 1d pipeline 
measures the radial velocity, stellar parameters, and elemental abundances of stars.
As of this writing, we routinely deliver the LAM 1d products to the user to further
enhance the scientific value of the PFS data. The GA 1d pipeline results are yet to be
included in the releases.

The pipeline outputs are split by proposal IDs and delivered to the PIs (i.e.,
a PI can only see spectra of their own targets). The processed data are served at
the PFS SP\footnote{\url{https://hscpfs.mtk.nao.ac.jp/}}. It offers
a cloud-based data analysis environment, where science users can analyze PFS data
without copying the data to their local computers over the network (it is
impractical to copy the massive data sets). Jupyter notebooks and remote desktop
environment can be used for the remote data analyses. Users can also ssh to
the system, mount the remote file system, and so on, all of which are useful for
intensive data mining.
In addition to the spectroscopic data from PFS, we also serve imaging data from
HSC PDR3 \citep{aihara22}. All of the HSC-SSP tools such as image browser, image cutout tool,
and catalog database are available, allowing the user to interactively analyze both
imaging and spectroscopic data on the same platform.

PFS SP utilizes the same access control system as STARS. STARS uses 
Lightweight Directory Access Protocol (LDAP) for managing user accounts and for maintaining a list of proposals that a user is part of.
The pipeline products are sorted into different directories for each proposal IDs,
and a user can access proposals for which the user is part of.
After the proprietary period is over, the entire proposal directory will be accessible
to everyone.
The only exception to this rule is the observatory filler; it does not have a proprietary
period, and anyone can access the observatory filler data. 

As mentioned earlier, we leave room for pipeline processing by the user. We provide
the user with the calibration products used in our processing as well as custom {\tt pfsConfig}
files at PFS SP. One needs to fiddle with
the pipeline configurations to build good calibrations, which is difficult for the general user.
We therefore deliver the calibrations to the user.
The pipeline tutorial describes how to import the calibration files and ingest {\tt pfsConfigs},
so that the user can start their processing without trouble.
The help desk will respond to any inquiries about the pipeline processing from the users.

%--------------------------------------------------------------------------------------------------------
\section{Summary and Future Improvements}
\label{sec:summary_and_future_improvements}

We have presented the new framework of the PFS science operations.
Due to the high multiplexity of the instrument over a wide field, we do not expect all PIs will
use all fibers efficiently, which means that we cannot efficiently operate the instrument
in Subaru’s current science operation framework.
That was the starting point of the new framework.
In order to fully exploit the capability of PFS, we have introduced a number of new
concepts such as fiber-share scheme, fiber hours (FHs), required observing time (ROT),
and effective exposure time (EET),
as well as new observing program categories such as community and
observatory fillers. New tools (e.g., target uploader, online ETC) have also been
developed to make the actual science operations possible. We emphasize that
the new framework is based on discussions with the Subaru community
over 4-5 years. The observatory has carefully listened to the community,
and the community input was an essential piece of the new framework.

There are five observing categories for PFS: normal queue, normal classical, ToO,
community filler, and observatory filler. 
The observatory filler is prepared by the observatory, but the remainder are by the community.
Proposals are selected through the standard competitive process, and priority and
data proprietary period are defined. The community filler category is rather inclusive
and we have introduced a distributed review system for the first time at the Subaru Telescope.
There is already a scientific result based on community filler data \citep{zhong26}.
The observatory filler is the lowest priority and has no proprietary period. It is a simple
magnitude-limited sample with no other selection functions, and we hope the community will
explore the observatory filler as well to carry out science.
Queue and filler programs are executed in queue mode and their relative priorities are
summarized in Table~\ref{tab:categories}. Classical programs as well as ToO
are executed in classical mode as of this writing.

The entire workflow is summarized in Fig. \ref{fig:flow}. The PI estimates
the exposure time using the ETC, in which the observing conditions are fixed to
the fiducial values, thereby giving the EET. The PI then uploads the target list
through the uploader, and includes the upload ID and ROT in the proposal. During
the TAC process, ROT is used to estimate the observing time of each proposal, and
FH is used for the actual time allocation. There is no phase 2 for PFS. For each
observing run, fiber designs for all programs are prepared by the observatory to
ensure that sufficient number of calibration objects are uniformly distributed across the field of view.
Data are reduced on site using the dedicated pipeline and EETs
are evaluated for each exposure and at each reference wavelength.  The EETs are
then used to update the progress of each program, feeding back to the observing
plan for the next night. After each observing run, all data are processed and
the reduced data are delivered to the PIs through PFS SP.

PFS has just started. Not all requests from the community have been implemented
in the real operation at this time. We have encountered new problems
in the first semester of the PFS operation.

As we have mentioned throughout the paper,
there is a lot of room for improvements. Just to list a few:

\begin{itemize}
\item Improve the simulator (which is a part of the online uploader) to run fast even when a large number of objects are uploaded.
\item Accept ToO programs in units of hours instead of half-nights.
\item Allow target upload/change for transient programs closer to observation.
\item Allow target updates for community filler programs during the semesters.
\end{itemize}

\noindent
Although not part of the core science operations, we also intend to improve
the data reduction pipelines to enhance the scientific output.
These possible improvements are not just technical issues, but also affect observatory policy.
PFS is new but it is not wise to deviate too much from the existing 
observatory-wide policies on subjects such as ToO. Careful considerations and coordination
are needed at the observatory, but we are confident that we will be able to make improvements.

We continue to collect input from the community for future improvements just like
we have done thus far to build the new  framework. This is important because the goal of the science
operations is to maximize the science output. Only when both the observatory and
community work together can such efficient and productive observations be possible.

%--------------------------------------------------------------------------------------------------------
\begin{ack}
  We dedicate the paper to the memory of Olivier LeFevre, Naruhisa Takato, and Hiroshige Yoshida,
  whose contributions were indispensable for the construction and science operations of PFS.
  
  The instrument \onohiula\ - Prime Focus Spectrograph (PFS) including both hardware
  and software was developed by the PFS collaboration to which over 25 institutes across
  multiple countries were committed. The technical activities were conducted by (in
  alphabetical order) Academia Sinica Institute of Astronomy and Astrophysics (Taiwan),
  California Institute of Technology, Johns Hopkins University, Kavli Institute for the
  Physics and Mathematics of the Universe in the University of Tokyo (Kavli IPMU),
  Laboratoire d'Astrophysique de Marseille, Laboratório Nacional de Astrofísica (Brazil),
  Max-Planck-Institut f\"{u}r Astrophysik, Max-Planck-Institut für extraterrestrische Physik,
  NASA Jet Propulsion Laboratory, National Astronomical Observatory of Japan (NAOJ),
  Princeton University, and Universidade de São Paulo, under the oversight by the Project
  Office hosted by Kavli IPMU (later NAOJ). There were also essential commitments from
  academic and industrial partners such as Durham University (United Kingdom) and and
  Bertin Technologies (France).
  
  The \onohiula\ PFS development work was supported by World Premier International
  Research Center Initiative (WPI), Ministry of Education, Culture, Sports, Science and
  Technology (MEXT), Japan. Kavli IPMU was established and supported by World
  Premier International Research Center Initiative (WPI), MEXT, Japan.
  We gratefully acknowledge support from the Funding Program for World-Leading
  Innovative R\&D on Science and Technology (FIRST) program ``Subaru Measurements of
  Images and Redshifts (SuMIRe)'' by Council for Science and Technology Policy (CSTP),
  Japan.

  \onohiula\ PFS makes use of the mechanical housing (so-called POpt2) that
  accommodates the Prime Focus Instrument and integrates the Wide Field Corrector
  lens system generating a flat focal plane with good image qualities across the wide field
  of view at the Subaru's prime focus. We appreciate all efforts to make these crucial
  components of infrastructure operational in conjunction with the development of Hyper
  Suprime-Cam (HSC).
  
  \onohiula\ PFS software components for instrument control and data processing utilize
  the platform developed and maintained for Vera C. Rubin Observatory. We appreciate
  their generosity of making it publicly available as an open source.
  We also appreciate the public catalogues from HSC-SSP PDR3, Gaia DR3, and the
  Pan-STARRS1 Surveys (PS1) which are exploited in PFS observations for field
  acquisition and auto-guiding of telescope pointing, characterization of sky spectra during
  exposures, and flux calibrations.

  This work is based on data collected at the Subaru Telescope, which is
  operated by NAOJ. We are honored and grateful for the opportunity of observing the
  Universe from Maunakea, which has the cultural, historical, and natural significance in
  Hawaii.

  The data analyses presented in this paper were carried out at the Prime Focus Spectrograph
  Science Platform, which is operated by the Subaru Telescope and the Astronomy Data Center
  at the National Astronomical Observatory of Japan.

  The HSC collaboration includes the astronomical communities of Japan and
  Taiwan, and Princeton University. The HSC instrumentation and software
  were developed by the National Astronomical Observatory of Japan (NAOJ),
  the Kavli Institute for the Physics and Mathematics of the Universe (Kavli
  IPMU), the University of Tokyo, the High Energy Accelerator Research
  Organization (KEK), the Academia Sinica Institute for Astronomy and
  Astrophysics in Taiwan (ASIAA), and Princeton University. Funding was
  contributed by the FIRST program from the Japanese Cabinet Office, the
  Ministry of Education, Culture, Sports, Science and Technology (MEXT), the
  Japan Society for the Promotion of Science (JSPS), Japan Science and
  Technology Agency (JST), the Toray Science Foundation, NAOJ, Kavli IPMU,
  KEK, ASIAA, and Princeton University.

  This work has made use of data from the European Space Agency (ESA)
  mission Gaia (https://www.cosmos.esa.int/gaia), processed by the Gaia Data
  Processing and Analysis Consortium (DPAC,
  https://www.cosmos.esa.int/web/gaia/dpac/consortium). Funding for the
  DPAC has been provided by national institutions, in particular the institutions
  participating in the Gaia Multilateral Agreement.

  The Pan-STARRS1 Surveys (PS1) and the PS1 public science archive have
  been made possible through contributions by the Institute for Astronomy, the
  University of Hawaii, the Pan-STARRS Project Office, the Max Planck
  Society and its participating institutes, the Max Planck Institute for
  Astronomy, Heidelberg, and the Max Planck Institute for Extraterrestrial
  Physics, Garching, The Johns Hopkins University, Durham University, the
  University of Edinburgh, the Queen’s University Belfast, the
  Harvard-Smithsonian Center for Astrophysics, the Las Cumbres Observatory
  Global Telescope Network Incorporated, the National Central University of
  Taiwan, the Space Telescope Science Institute, the National Aeronautics and
  Space Administration under grant No. NNX08AR22G issued through the
  Planetary Science Division of the NASA Science Mission Directorate, the
  National Science Foundation grant No. AST-1238877, the University of
  Maryland, Eotvos Lorand University (ELTE), the Los Alamos National
  Laboratory, and the Gordon and Betty Moore Foundation.
\end{ack}

\section*{Funding}
 This work is supported by Japan Society for the Promotion of Science (JSPS)
  KAKENHI Grant Numbers JP15H05893, JP15K21733, JP15H05892, JP20H05850,
  JP20H05855, and JP23H05438.
  The work at Princeton University, Johns Hopkins University, and California Institute of
  Technology is supported in part by NSF Award 1636426.
  The work in ASIAA, Taiwan, is supported by the Academia Sinica of Taiwan.
  The work in Brazil is supported by grants from CNPq (308994/2021-3) and FAPESP
  (2011/51680-6).
  The work in France is supported by CNRS and Aix Marseille University.

%\section*{Data availability} 
% The data underlying this article are available ...  
% Sample Data Availability Statements 
% https://academic.oup.com/pages/open-research/research-data#Data%20Availability%20Statements

%\appendix %%%%%%%%%%%%%%%%%%%%%%%%%%%%%%%%%%%%%%%%%%%%%%%%%%%%%%%%
%\section*{Case of single paragraph}
% No section number is necessary. Add ``*'' after \verb/\section/.
%
%\section{Case of two or more paragraphs}
%
%\section{Case of two or more paragraphs}
%

%--------------------------------------------------------------------------------------------------------
% Any journal's BST file (e.g., apj.bst) can be used as PASJ's BST is unavailable.    


\begin{thebibliography}{}
\bibitem[Aihara et al.(2018a)]{aihara18a} Aihara, H., Arimoto, N., Armstrong, R., et al.\ 2018, \pasj, 70, S4. doi:10.1093/pasj/psx066
\bibitem[Aihara et al.(2018b)]{aihara18b} Aihara, H., Armstrong, R., Bickerton, S., et al.\ 2018, \pasj, 70, S8. doi:10.1093/pasj/psx081
\bibitem[Aihara et al.(2019)]{aihara19} Aihara, H., AlSayyad, Y., Ando, M., et al.\ 2019, \pasj, 71, 114. doi:10.1093/pasj/psz103
\bibitem[Aihara et al.(2022)]{aihara22} Aihara, H., AlSayyad, Y., Ando, M., et al.\ 2022, \pasj, 74, 247. doi:10.1093/pasj/psab122
\bibitem[Akeson et al.(2019)]{akeson19} Akeson, R., Armus, L., Bachelet, E., et al.\ 2019, , arXiv:1902.05569. doi:10.48550/arXiv.1902.05569
\bibitem[Andrae et al.(2023)]{andrae23} Andrae, R., Fouesneau, M., Sordo, R., et al.\ 2023, \aap, 674, A27. doi:10.1051/0004-6361/202243462
\bibitem[Chambers et al.(2016)]{chambers16} Chambers, K.~C., Magnier, E.~A., Metcalfe, N., et al.\ 2016, , arXiv:1612.05560. doi:10.48550/arXiv.1612.05560
\bibitem[Chiba et al.(2026)]{chiba26} Chiba, M., Wyse, R.~F.~G., Kirby, E.~N., et al.\ 2026, arXiv:2604.09875. doi:10.48550/arXiv.2604.09875
\bibitem[Cirasuolo et al.(2020)]{cirasuolo20} Cirasuolo, M., Fairley, A., Rees, P., et al.\ 2020, The Messenger, 180, 10. doi:10.18727/0722-6691/5195
\bibitem[Coupon et al.(2018)]{coupon18} Coupon, J., Czakon, N., Bosch, J., et al.\ 2018, \pasj, 70, S7. doi:10.1093/pasj/psx047
\bibitem[de Jong et al.(2019)]{dejong19} de Jong, R.~S., Agertz, O., Berbel, A.~A., et al.\ 2019, The Messenger, 175, 3. doi:10.18727/0722-6691/5117
\bibitem[de Oliveira et al.(2022)]{deoliveira22} de Oliveira, A.~C., de Oliveira, L.~S., Ferreira, D., et al.\ 2022, \procspie, 12184, 1218474. doi:10.1117/12.2629555
\bibitem[DESI Collaboration et al.(2016a)]{desi16a} DESI Collaboration, Aghamousa, A., Aguilar, J., et al.\ 2016, , arXiv:1611.00036. doi:10.48550/arXiv.1611.00036
\bibitem[DESI Collaboration et al.(2016b)]{desi16b} DESI Collaboration, Aghamousa, A., Aguilar, J., et al.\ 2016, , arXiv:1611.00037. doi:10.48550/arXiv.1611.00037
\bibitem[Euclid Collaboration et al.(2025)]{euclid25} Euclid Collaboration, Mellier, Y., Abdurro'uf, et al.\ 2025, \aap, 697, A1. doi:10.1051/0004-6361/202450810
\bibitem[Flewelling et al.(2020)]{flewelling20} Flewelling, H.~A., Magnier, E.~A., Chambers, K.~C., et al.\ 2020, \apjs, 251, 1, 7. doi:10.3847/1538-4365/abb82d
\bibitem[Gaia Collaboration et al.(2016)]{gaia16} Gaia Collaboration, Prusti, T., de Bruijne, J.~H.~J., et al.\ 2016, \aap, 595, A1. doi:10.1051/0004-6361/201629272
\bibitem[Gaia Collaboration et al.(2023)]{gaia23} Gaia Collaboration, Vallenari, A., Brown, A.~G.~A., et al.\ 2023, \aap, 674, A1. doi:10.1051/0004-6361/202243940
\bibitem[Greene et al.(2022)]{greene22} Greene, J., Bezanson, R., Ouchi, M., et al.\ 2022, arXiv:2206.14908. doi:10.48550/arXiv.2206.14908
\bibitem[Gunn et al.(1998)]{gunn98} Gunn, J.~E., Carr, M., Rockosi, C., et al.\ 1998, \aj, 116, 6, 3040. doi:10.1086/300645
\bibitem[Gunn et al.(2006)]{gunn06} Gunn, J.~E., Siegmund, W.~A., Mannery, E.~J., et al.\ 2006, \aj, 131, 4, 2332. doi:10.1086/500975
\bibitem[Guy et al.(2023)]{guy23} Guy, J., Bailey, S., Kremin, A., et al.\ 2023, \aj, 165, 4, 144. doi:10.3847/1538-3881/acb212
\bibitem[Hanuschik(2003)]{hanuschik03} Hanuschik R.~W., 2003, A\&A, 407, 1157. doi:10.1051/0004-6361:20030885
\bibitem[Hinkle et al.(2003)]{hinkle03} Hinkle, K.~H., Wallace, L., \& Livingston, W.\ 2003, AAS, 203, 38.03. 
\bibitem[Ivezi{\'c} et al.(2019)]{ivezic19} Ivezi{\'c}, {\v{Z}}., Kahn, S.~M., Tyson, J.~A., et al.\ 2019, \apj, 873, 2, 111. doi:10.3847/1538-4357/ab042c
\bibitem[Iye et al.(2004)]{iye04} Iye, M., Karoji, H., Ando, H., et al.\ 2004, \pasj, 56, 381. doi:10.1093/pasj/56.2.381
\bibitem[Krisciunas \& Schaefer(1991)]{krisciunas91} Krisciunas K. \& Schaefer B.~E.\ 1991, \pasp, 103, 1033. doi:10.1086/132921
\bibitem[Lawrence et al.(2007)]{lawrence07} Lawrence, A., Warren, S.~J., Almaini, O., et al.\ 2007, \mnras, 379, 4, 1599. doi:10.1111/j.1365-2966.2007.12040.x
\bibitem[Lord (1992)]{lord92} Lord, S. D., 1992, NASA Technical Memorandum 103957
\bibitem[Magnier et al.(2020)]{magnier20} Magnier, E.~A., Schlafly, E.~F., Finkbeiner, D.~P., et al.\ 2020, \apjs, 251, 1, 6. doi:10.3847/1538-4365/abb82a
\bibitem[Miyazaki et al.(2018)]{miyazaki18} Miyazaki, S., Komiyama, Y., Kawanomoto, S., et al.\ 2018, \pasj, 70, S1. doi:10.1093/pasj/psx063
\bibitem[Oke \& Gunn(1983)]{oke83} Oke, J.~B. \& Gunn, J.~E.\ 1983, \apj, 266, 713. doi:10.1086/160817
\bibitem[Rousselot et al.(2000)]{rousselot00} Rousselot P., Lidman C., Cuby J.-G., Moreels G., Monnet G., 2000, A\&A, 354, 1134.
\bibitem[Schmitt et al.(2019)]{schmitt19} Schmitt, A., Arnouts, S., Borges, R., et al.\ 2019, Astronomical Data Analysis Software and Systems XXVI, 521, 398.
\bibitem[Scoville et al.(2007)]{scoville07} Scoville, N., Aussel, H., Brusa, M., et al.\ 2007, \apjs, 172, 1, 1. doi:10.1086/516585
\bibitem[Smee et al.(2022)]{smee22} Smee, S.~A., Gunn, J.~E., Barkhouser, R.~H., et al.\ 2022, \procspie, 12184, 121847L. doi:10.1117/12.2630536
\bibitem[Speagle et al.(2025)]{speagle25} Speagle, J.~S., Zucker, C., Beane, A., et al.\ 2025, arXiv:2503.02227. doi:10.48550/arXiv.2503.02227
\bibitem[Tachibana \& Miller(2018)]{tachibana18} Tachibana, Y. \& Miller, A.~A.\ 2018, \pasp, 130, 994, 128001. doi:10.1088/1538-3873/aae3d9
\bibitem[Takada et al.(2014)]{takada14} Takada, M., Ellis, R.~S., Chiba, M., et al.\ 2014, \pasj, 66, 1, R1. doi:10.1093/pasj/pst019
\bibitem[Tamura et al.(2022)]{tamura22} Tamura, N., Moritani, Y., Yabe, K., et al.\ 2022, \procspie, 12184, 1218410. doi:10.1117/12.2628152
\bibitem[Tamura et al.(2024)]{tamura24} Tamura, N., Yabe, K., Koshida, S., et al.\ 2024, \procspie, 13096, 1309605. doi:10.1117/12.3015967
\bibitem[Tanaka et al.(2017)]{tanaka17} Tanaka, M., Hasinger, G., Silverman, J.~D., et al.\ 2017, arXiv:1706.00566. doi:10.48550/arXiv.1706.00566
\bibitem[Tulloch et al.(2019)]{tulloch19} Tulloch, S., George, E., \& ESO Detector Systems Group\ 2019, Journal of Astronomical Telescopes, Instruments, and Systems, 5, 036004. doi:10.1117/1.JATIS.5.3.036004
\bibitem[Winegar(2008)]{winegar08} Winegar, T.\ 2008, \procspie, 7016, 70160M. doi:10.1117/12.789832
\bibitem[Wang et al.(2020)]{wang20} Wang, S.-Y., Chou, C.-Y., Chang, Y.-C., et al.\ 2020, \procspie, 11447, 1144784. doi:10.1117/12.2561191
\bibitem[Wang et al.(2022)]{wang22} Wang, S.-Y., Kimura, M., Yan, C.-H., et al.\ 2022, \procspie, 12184, 121846R. doi:10.1117/12.2629098
\bibitem[York et al.(2000)]{york00} York, D.~G., Adelman, J., Anderson, J.~E., et al.\ 2000, \aj, 120, 3, 1579. doi:10.1086/301513
\bibitem[Zhang et al.(2025)]{zhang25} Zhang, H., Kokubo, M., MacBride, S., et al.\ 2025, , arXiv:2508.00291. doi:10.48550/arXiv.2508.00291
\bibitem[Zhong et al.(2026)]{zhong26} Zhong, Y., Chen, X., Ichikawa, K., et al.\ 2026, arXiv:2603.13736. doi:10.48550/arXiv.2603.13736
\end{thebibliography}
\end{document}